\documentclass[traditabstract]{aa} 
\usepackage{graphicx}
\usepackage{txfonts}
\begin{document}
   \title{The spectroscopic evolution of the $\gamma$-ray emitting classical nova Nova Mon 2012}     
   \subtitle{I. Implications for the ONe subclass of classical novae}
   \author{S. N. Shore\inst{1,2}, I. De Gennaro Aquino\inst{1}, G. J. Schwarz\inst{3},\\ T. Augusteijn\inst{4},
     C. C. Cheung\inst{5}, F. M. Walter\inst{6}, and S. Starrfield\inst{7}}
        \institute{
 Dipartimento di Fisica ``Enrico Fermi'', Universit\`a di Pisa, and INFN- Sezione Pisa, largo
B. Pontecorvo 3, I-56127 Pisa, Italy; \email{shore@df.unipi.it,ivan.degennaroaquino@gmail.com}
\and
INFN - Sezione Pisa, largo B. Pontecorvo 3, I-56127 Pisa, Italy
  \and
  American Astronomical Society, 2000 Florida Ave NW, Washington DC 20009-1231, USA; \email{Greg.Schwarz@aas.org}
    \and
  Nordic Optical Telescope, Apartado 474, E-38700 Santa Cruz de La Palma,
Santa Cruz de Tenerife,Spain; \email{tau@not.iac.es}
 \and
 Space Science Division, Naval Research Laboratory, Washington, DC 20375-5352, USA; \email{Teddy.Cheung@nrl.navy.mil}
 \and
 Department of Physics and Astronomy, Stony Brook University Stony Brook NY 11794-3800, USA; \email{frederick.walter@stonybrook.edu}
 \and
 School of Earth and Space Exploration,  Arizona State University, P.O. Box 871404, Tempe, AZ 85287-1404,USA  \email{sumner.starrfield@asu.edu} }

              \date{received ---; accepted ---}


 \abstract
 {
   Among the classical novae, the ONe subgroup, distinguished by their large overabundance of neon, are thought to occur on the most massive white dwarfs.  Nova Mon 2012 was the first classical nova  to be detected as a high energy $\gamma$-ray transient, by {\it Fermi}-LAT, before its optical discovery.  The first optical spectra obtained about 55 days after $\gamma$-ray peak, were strikingly similar to the ONe class after the transition to the nebular (optically thin) spectrum.  The current paper presents our subsequent optical and ultraviolet observations.   A time sequence of optical echelle spectra (3700-7400\AA) with the Nordic Optical Telescope (NOT) began on 2012 Aug. 16, immediately following the optical announcement, and are continuing.  The nova was observed almost simultaneously with the NOT on 2012 Nov. 21, with the Space Telescope Imaging Spectrograph (STIS) aboard the Hubble Space Telescope at medium echelle resolution (1150-3050\AA) on Nov. 20, and  with the CHIRON CTIO/SMARTS echelle spectrograph at medium resolution (4500-8900\AA) on Nov. 22.  We use plasma diagnostics (e.g. [O III] and H$\beta$ line flux) to constrain electron densities and temperatures, and the filling factor, for the ejecta.  Using Monte Carlo modeling, we derive the structure from comparisons to the optical and ultraviolet line profiles.  We also compare observed fluxes for Nova Mon 2012 with those predicted by photoionization modeling with Cloudy using element abundances derived for other ONe novae, the parameters derived from the line profile modeling and multiwavelength continuum measurements.  Nova Mon 2012 is confirmed as an ONe nova first observed spectroscopically in the nebular stage.  We derive an extinction of E(B-V)=0.85$\pm$0.05 and hydrogen column density $N_H \approx 5\times 10^{21}$ cm$^{-2}$.  The corrected continuum luminosity is nearly the same in the entire observed energy range compared to V1974 Cyg, V382 Mon, and Nova LMC 2000 at the same epoch after outburst.   The distance, about 3.6 kpc, is quite similar to V1974 Cyg, suggesting that it would have been equally bright had it been observed at maximum light.  The same applies to the line profiles.  These can be modeled using an  axisymmetric conical -- bipolar -- geometry for the ejecta with various inclinations of the axis to the line of sight, $i$, and ejecta inner radii.  For Nova Mon 2012, we find that  $60 \le i \le 80$ degrees, an opening angle of $\approx$70$^o$, and an inner radius $\Delta R/R(t)\approx 0.4$ matches the permitted and intercombination lines  while the forbidden lines require a less filled structure.  The filling factor is $f\approx 0.1-0.3$, although it may be lower based on the structures observed in the emission line profiles,  implying an ejecta mass $\leq 6\times 10^{-5}$M$_\odot$.  The abundances are similar to, but not identical to, V1974 Cyg and V382 Vel.  In particular, Ne and Mg are apparently more abundant in Nova Mon 2012.  The ONe novae appear to comprise a single physical class with bipolar high mass ejecta, similarly enhanced abundances, and a common spectroscopic evolution within a narrow range of luminosities.   The spectral evolution does not require continued mass loss from the post-explosion white dwarf.  This also implies that the detected $\gamma$-ray emission is a generic phenomenon, common to all ONe novae, possibly to all classical novae, and connected with acceleration and emission  processes  within the ejecta. }


          \thanks{Based on observations made with the NASA/ESA Hubble Space Telescope, obtained from the data archive at the Space Telescope Science Institute. STScI is operated by the Association of Universities for Research in Astronomy, Inc. under NASA contract NAS 5-26555.}
          \thanks{Based on observations made with the Nordic Optical Telescope, operated on the island of La Palma jointly by Denmark, Finland, Iceland, Norway, and Sweden, in the Spanish Observatorio del Roque de los Muchachos of the Instituto de Astrofisica de Canarias. }

  \titlerunning{Nova Mon 2012 and the unification of the ONe novae. I.} \authorrunning{S. N.
Shore et al.}

 \maketitle

\section{Introduction}

Nova Mon 2012  was the first classical nova detected at high energies ($>$ 100 MeV) {\it before} optical discovery.  It was reported as  a $\gamma$-ray transient {\it Fermi}-LAT source, designated J0639+0548, on 2012 Jun. 29 (Cheung et al. 2012a) based on data from Jun. 18.0-26.4 (MJD 56096.0 - 104.4).  The first optical report was 2012 August 9 by S. Fujikawa (CBET 3202) with visual magnitude 9.4 and position $\alpha_{J2000} = 6^h39^m38.^s57$ and $\delta_{J2000} = +5^\circ 53'53".4$, designated PNV J06393874+0553520.   Follow-up spectroscopic observations (Cheung et al. 2012b) confirmed that this source was a classical nova observed after the beginning of the nebular stage, hence considerably after the optically thick period of maximum light.  This agreed with the identification of the $\gamma$-ray transient as the onset of the explosion.   The earliest optical low resolution spectra obtained by the ARAS group\footnote{http://www.astrosurf.com/aras/novae/Nova2012Mon.html} and high resolution spectra obtained with the Nordic Optical Telescope (see below) showed a striking resemblance to a similar stage in V382 Vel (Della Valle et al. 2002),  one of the brightest novae of the 20th century and an ONe nova (see Fig. 1).  This subclass of classical novae is distinguished by a very substantial overabundance of oxygen, neon, and magnesium.  These novae are thought to originate via an accretion-induced  thermonuclear runaway (TNR)  and deep envelope mixing on a near-Chandrasekhar mass white dwarf (WD) (see e.g. Starrfield et al. (2008) and Downen et al. (2012)).   During the months since optical announcement, we have been carrying out a multiwavelength campaign.  Here report the first results of the high resolution optical and ultraviolet spectroscopy.  We also realized, during the course of this analysis, that there is a developmental sequence and physical properties linking the four ONe novae observed in the last 20 years at high resolution from 1200-8000\AA\ that points to their possibly being a more unified group than previously thought, perhaps even a standard candle. 

\begin{figure}
\centerline{
\includegraphics[angle=0,width=10cm]{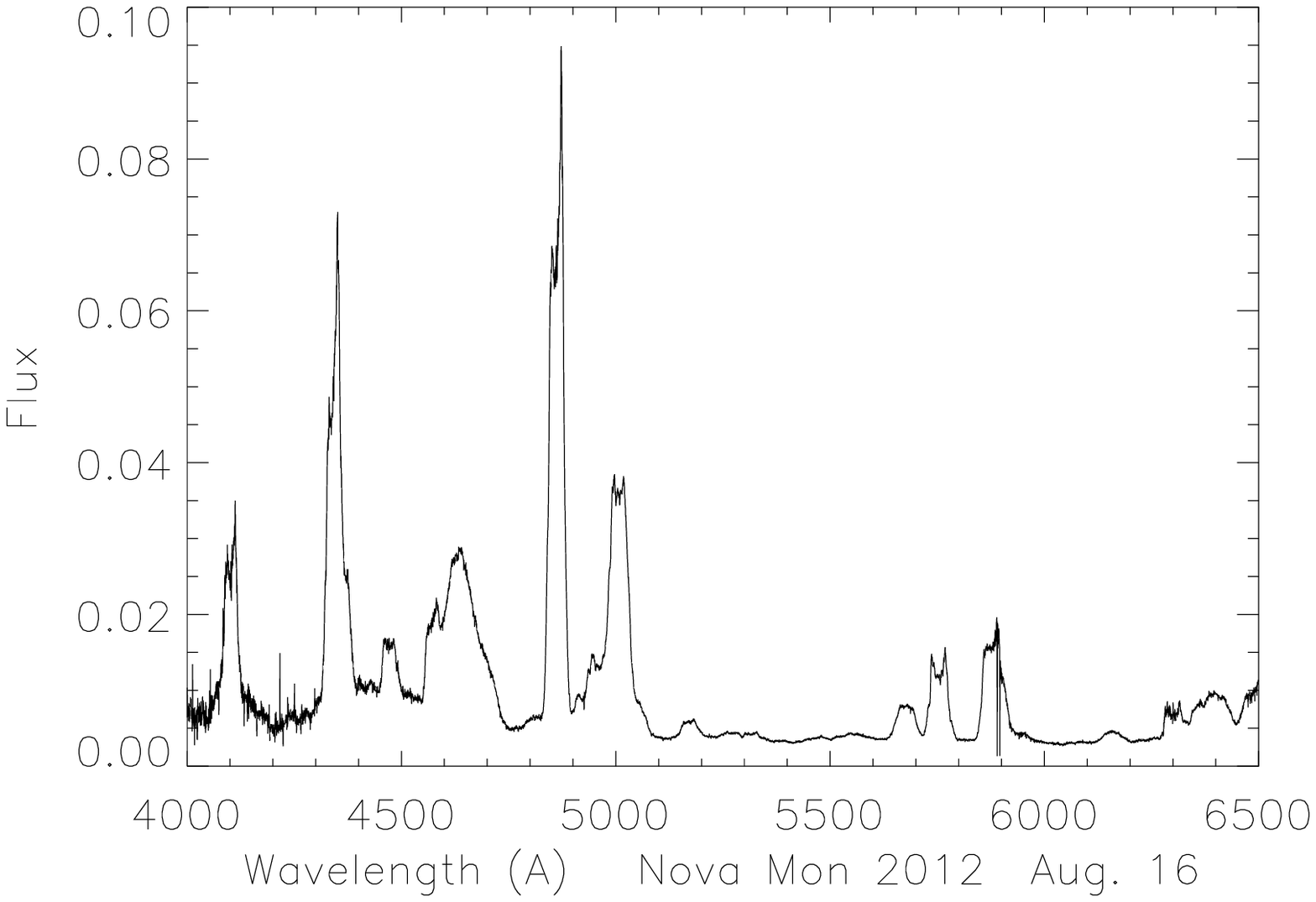}}

            \centerline{\includegraphics[width=9cm]{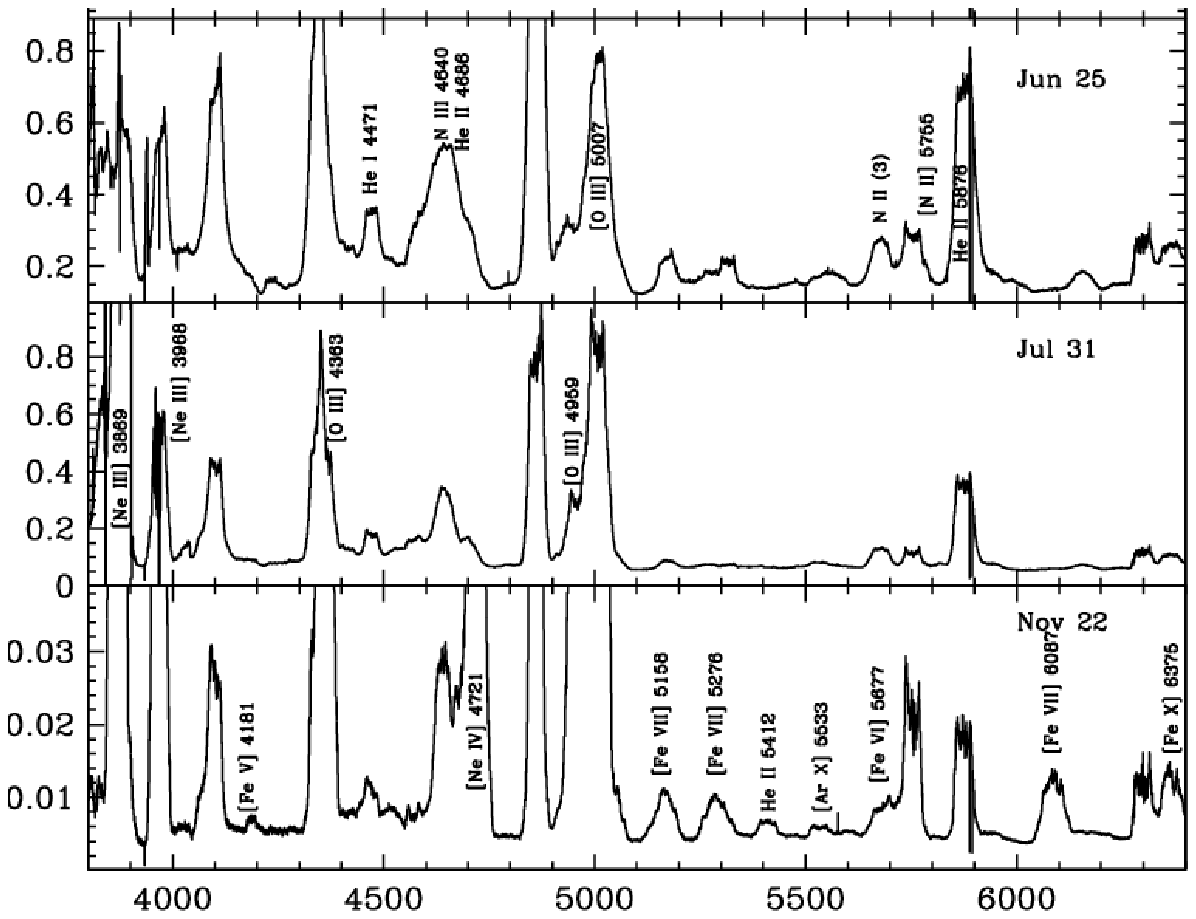}}

\caption{Top: First NOT spectrum of Nova Mon 2012, from Aug. 16.  Bottom: Late stage optical spectra from 1999 of V382 Vel from Della Valle et al. (2002) shown for comparison.  It was based on the comparison of these three spectra with the Aug. 16 observation that the {\it Fermi}-LAT $\gamma$-ray transient was identified with the peak of the outburst  (Cheung et al. 2012a,b). \label{f:many}}
\end{figure}

\section{Observational Data}

Our principal optical data set consists of spectra taken from 2012 Aug. 16 through Nov. 21 with the   2.6 m Nordic Optical Telescope
(NOT) FIber-fed Echelle Spectrograph (FIES)  with a dispersion of 0.023 \AA\ px$^{-1}$ in high-resolution mode, covering the spectral interval 3635 - 7364 \AA.   The first spectrum (Aug. 16) was not absolutely calibrated but the subsequent spectra were, using HD 289002 (= Hiltner 600) as the standard and checking the accuracy with low resolution contemporary spectra from SMARTS.\footnote{The archive of optical spectra is available at http://www.astro.sunysb.edu/fwalter/SMARTS/NovaAtlas/}   The CHIRON spectrum was obtained on 2012 Aug. 22 on the 1.5 meter at CTIO with a dispersion of 0.07\AA\ px$^{-1}$ covering the range 4500 - 8900\AA.  The journal of optical spectra is given in Table 1a.  All NOT spectra were reduced using IRAF, FIESTool, and IDL, the CHIRON spectrum was reduced with IRAF.\footnote{IRAF is   distributed by the National Optical Astronomy Observatories, which  are operated by the Association of Universities for Research in
  Astronomy, Inc., under cooperative agreement with the US National
  Science Foundation.}     Associated with these, we had a (fortuitously) simultaneous ultraviolet (UV) observation with the Space Telescope Imaging Spectrograph (STIS) aboard the {\it Hubble Space Telescope}.   We obtained medium resolution echelle spectra from 1150 - 3100\AA.   The journal of observations is given in Table 1b.  {\bf Unlike the conventional designation of epoch for nova outbursts, note that interval will hereafter be referred to as  the number of days after the initial peak of the $\gamma$-ray outburst on 2012 Jun. 22 (MJD 56100).}
  
\begin{center}
Table 1a: Journal of NOT/FIES and CHIRON observations \\
\begin{tabular}{llllr}
\hline
NOT &                     &           &                                        \\
\hline
Date  & Time (UT) & MJD  & Day & t$_{exp}$ (sec) \\
\hline
2012 Aug 16 & 05:49:40.7  & 56155.243 & 55 & 300  \\
2012 Sep. 6 & 05:05:39.3  & 56176.212 & 76 & 2000  \\
2012 Oct. 6 & 05:01:10.2    &  56206.049 & 106 & 3000 \\
2012 Nov. 21 & 01:26:08.4  & 56252.060 & 152 & 2765  \\
\hline
CHIRON &                            &                      &               \\
\hline
2012 Nov 22 & 06:10:00.4 & 56253.257 & 153 & 1800 \\
\hline
\end{tabular}
\end{center}

\begin{center}
Table 1b: Journal of {\it HST}/STIS observations \\
\begin{tabular}{llll}
\hline
STIS &  2012 Nov.  20  (GO 13120) &  Day 151 &  \\
\hline
OBSID &  MJD  & t$_{exp}$ (sec) & Grating \\
\hline
oc4701010 &  56251.896 & 900 & E140M \\
oc4701020 &  56251.912 & 310 & E230M  \\
oc4701020 &  56251.919  & 410 & E230M  \\
\hline
\end{tabular}
\end{center}

\section{Line profile evolution}

Since the optical data were obtained after the transition to a nebular spectrum, the line profiles showed relatively little overall evolution.  There were, however, several important changes.  Some were related to the expansion and others were caused by changes in the luminosity and spectrum of the relic WD.   We briefly comment on the individual results to accompany the display of the data.\footnote{The optical line identifications were greatly aided by the finding list provided as the appendix in Williams (2012).}  The next section will deal with the quantitative analyses.

\subsection{Balmer, He I, and He II lines}

From the first spectra, obtained on day 55, the Balmer lines showed a consistently asymmetric profile (Fig. 2).  This was quickly recognized to be similar to other novae in the transition stage and prompted the comparison with V382 Vel 1999 -- especially spectra from Della Valle et al. (2002) -- that was announced in the first report of the identification of Nova Mon 2012 with the {\it Fermi}-LAT transient  (Cheung et al.2012b).  The optical spectrum was quite similar to V382 Vel between day 50 and  day 90, an observation that corroborated the association of the transient with the onset of the event to within a few days.  The subsequent changes in the Balmer profiles were comparatively subtle through Nov. 21, the last NOT spectrum.  The main changes were the increasing contrast  between the emission feature at +1000 km s$^{-1}$ and the increasing prominence of the depression between -600 and 600 km s$^{-1}$.  The relative strength of the +1000 km s$^{-1}$ emission decreased passing from H$\alpha$ to H$\delta$.  The profiles were more symmetric for the higher members of the Balmer series.  We suggest that the changes in the individual peaks was due to recombination during the initial decline from Aug. through Oct. and then re-ionization of the ejecta following the turn-on of the X-ray supersoft source (SSS) (Nelson et al. (2012), see sec. 4 for further discussion).

  \begin{figure*}
   \centering
   \includegraphics[width=13cm,angle=90]{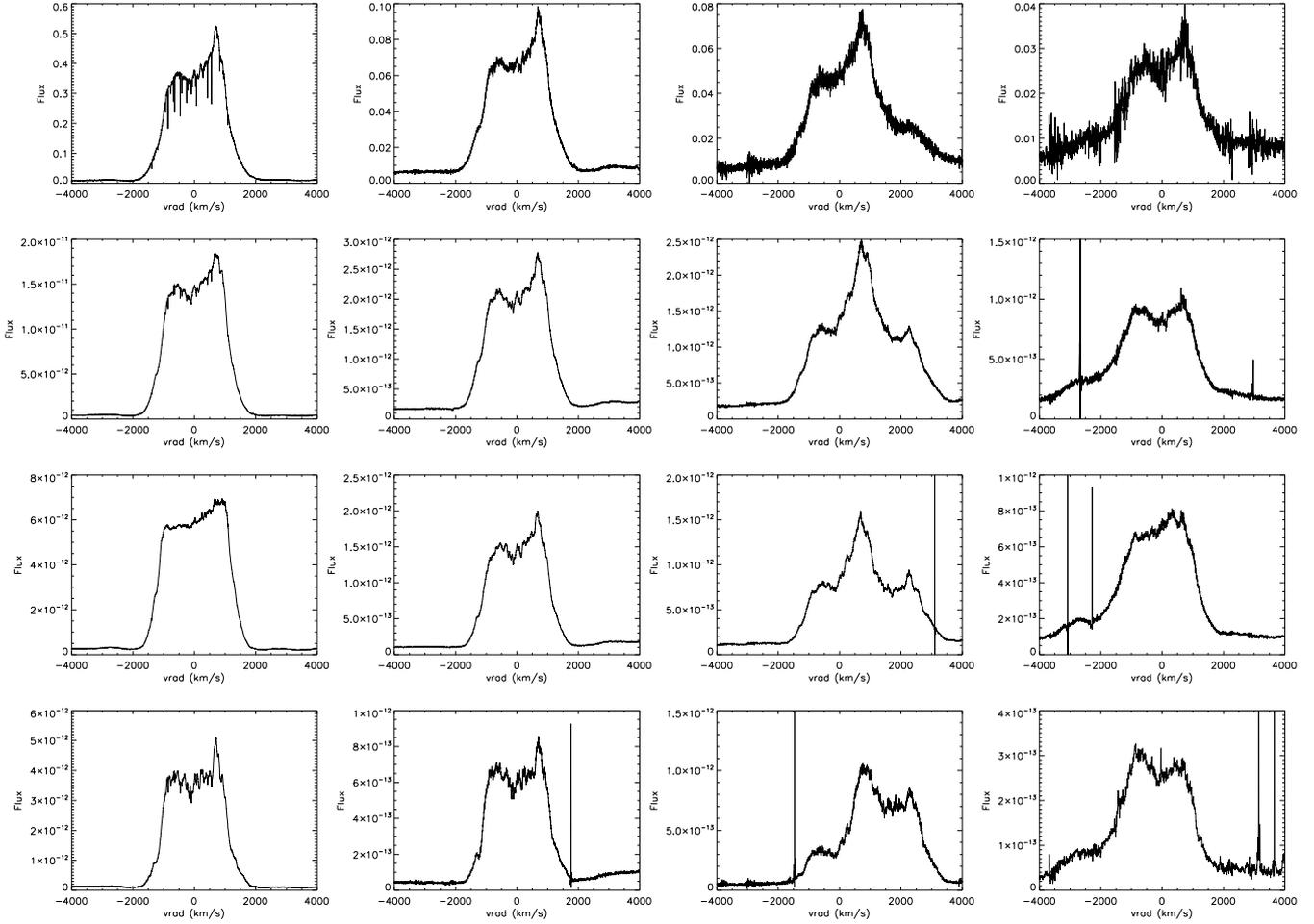}
   \caption{Gallery of H$\alpha$ (first column), H$\beta$ (second right),  H$\gamma$ (third, right), and H$\delta$ (far right) line profiles from the NOT observations.  From top to bottom:  Day 55, 76, 106, and 152.   Note the extreme blending between H$\gamma$ and [O III] 4363\AA\ throughout the sequence.  In addition, the [N II] 6548, 6583\AA\ lines were marginally visible in the wings of the Nov. 21 H$\alpha$ profile  The spikes are unremoved cosmic ray hits.}
    \end{figure*}

\begin{figure*}
   \centering
   \includegraphics[width=16cm]{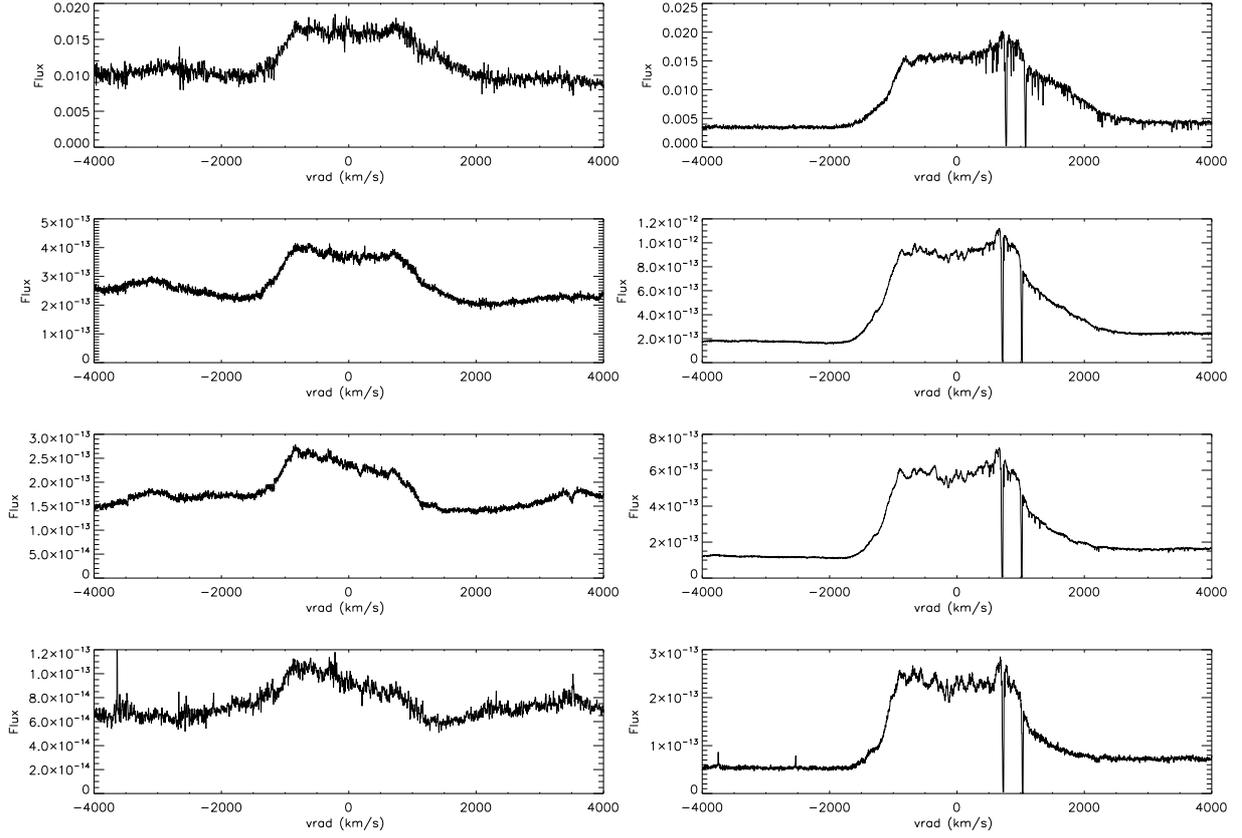}
   \caption{ Gallery of the He I 4471\AA\ (left) and 5876\AA\ (right) line profiles from the NOT spectra  on  Days 55, 76, 106, 152 .} 
 \end{figure*}

\begin{figure}
   \centering
   \includegraphics[width=9cm]{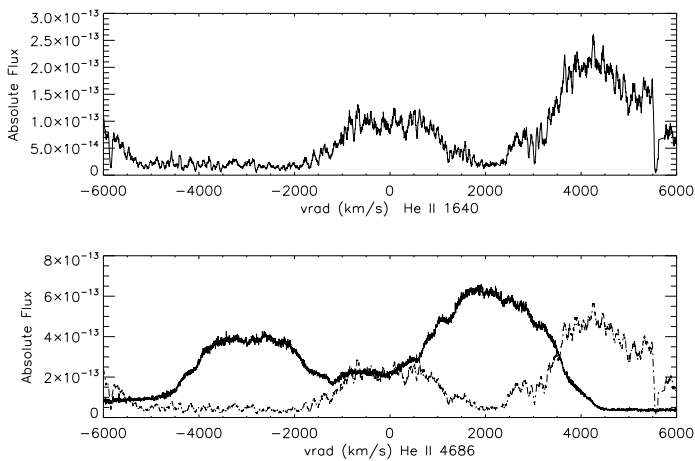}
   \caption{ Comparison of the He II 1640\AA\ (top: STIS spectrum) and 4868\AA\ (bottom: NOT spectrum) line profiles from Nov. 20/21.  Note that the two spectral intervals were nearly simultaneous, making this comparison possible .  The He II 4686\AA\ is severely blended but the line accounts for the excess emission.  In the bottom panel, the He II 1640\AA\ line is shown (dashed line) for comparison scaled by the continuum flux.}
 \end{figure}

The He I lines showed several different line profiles. The He I 6678\AA\ line -- the only singlet transition detected in the entire time series -- was always weak and double peaked (see Fig. 10).  He I 5876\AA\ and He I 7065\AA\ showed a similar profile to H$\alpha$ and H$\beta$ with the same emission peak at +1000 km s$^{-1}$.  Neither profile varied substantially during the observing interval.  In contrast, the He I 4471\AA\ line showed a distinct, strongly asymmetric profile that overall was more similar to the H$\delta$ line than the other Balmer profiles.  It also varied systematically in the same way as H$\delta$ during the time sequence (see Fig. 3).   

The He II 1640\AA\ line was easily detected in the STIS spectrum from Nov. 20.  Although weakly exposed, the profile resembled more closely the simultaneously observed H$\delta$ line than the other Balmer lines, lacking the strong emission at +1000 km s$^{-1}$ and having a less pronounced central depression.  He II 4686\AA\ was never sufficiently resolved from N III 4640\AA\ for a meaningful comparison although a superposition of the 1640\AA\ line as a proxy line profile suffices to explain the excess emission between the N III and He I 4713\AA\ (see fig. 4).  The wing structure was also different between the He$^\circ$ and He$^+$ lines (see sec. 4).

\subsection{Carbon, Nitrogen, Oxygen}

Three ionization stages were observed for carbon.  The UV resonance doublet C II 1334, 1335\AA\ was weak but present, and the C II] 2323, 2324\AA\ doublet showed a composite profile from the blend.  The C III] 1909\AA\ intercombination line was detected (although weakly exposed) with a profile cthat was similar to other intercombination and forbidden lines (especially [N II] 5755\AA).   The C IV 1548,1550\AA\ resonance doublet was strong and consistent with two blended profiles each of which were similar to H$\beta$.  The interstellar C IV doublet lines were easily measured (see sec. 4) since they were within the broader emission profile.

Four ionization stages of  nitrogen were observed, although only two were obtained in the optical time series.  The lowest ion, N$^+$, was observable only for [N II] 5755\AA\ (see Fig. 5).  The [N II] 6548, 6583\AA\ lines, that are the corresponding isoelectronic transitions to the nebular lines of  [O III], were always too severely blended in the wings of H$\alpha$ to be usable; they are likely barely visible in the Nov. 21 spectrum.  The N$^{+2}$-N$^{+4}$ ions were represented by N III]1750\AA\ (present although badly underexposed),  N III 4640\AA\ blend (see Fig. 4), N IV] 1486\AA, and N V 1238, 1242\AA.   We will concentrate on these few transitions since they provide the most important probes of the ejecta properties, but many higher excitation lines of N II and N III were present throughout the spectrum.

\begin{figure*}
   \centering
   \includegraphics[width=18cm]{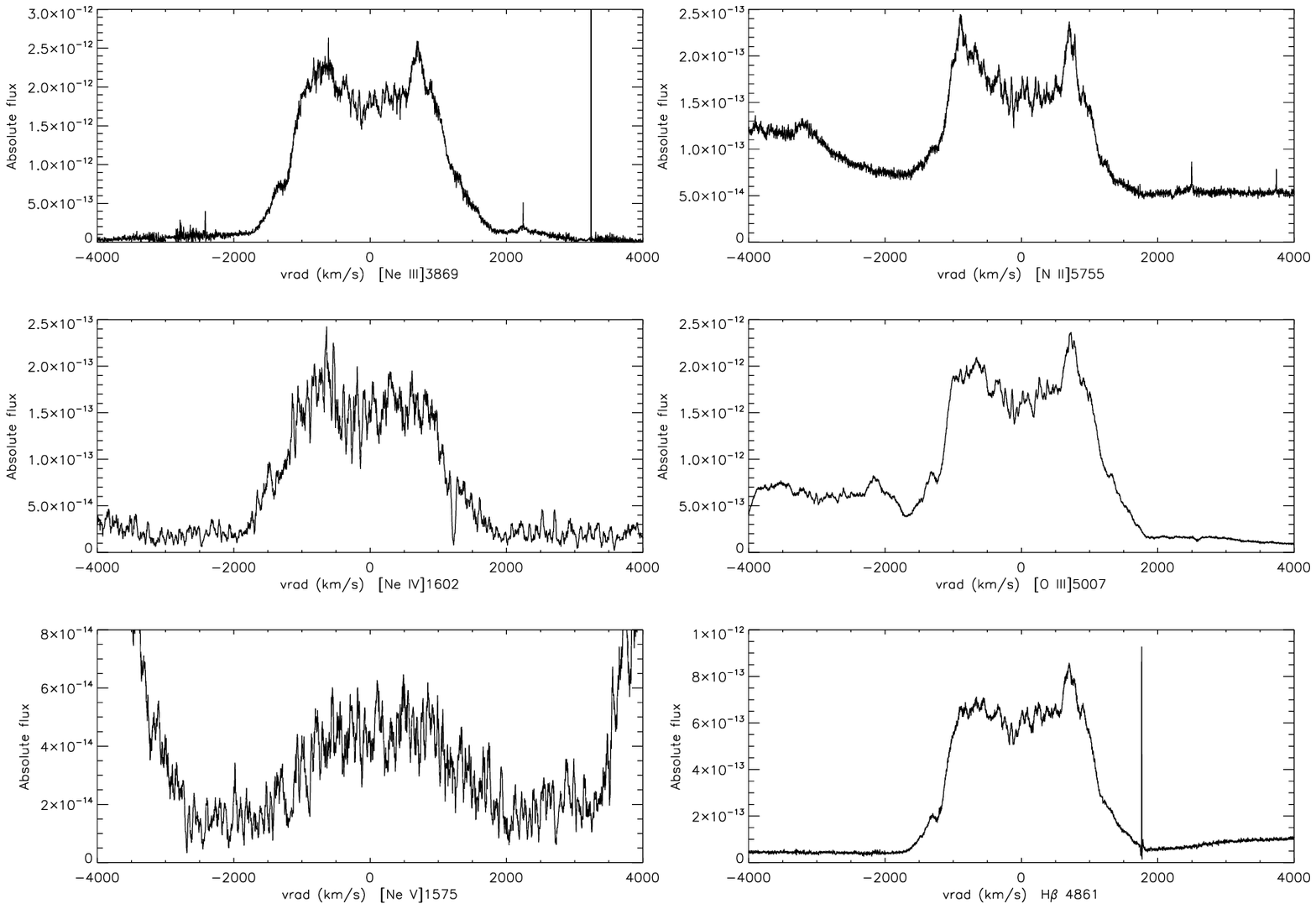}
   \caption{Comparisons of the [Ne III], Ne IV],  and [Ne V] (left column) with [N II] 5755\AA, [O III] 5077\AA\ and H$\beta$ (righ column)  from the Nov. NOT and STIS spectra.  See text for details.}
\end{figure*}

Similarly, four ionization stages were observed for oxygen.  The [O I] 6300\AA\ line, although blended with terrestrial bands, was detected throughout the sequence with a similar profile to [N II] 5755; in contrast, the [O I] 5577\AA\ line was never detected.   The O I 8446.3\AA\ line was present in the Nov. 22 CHIRON spectrum and is compared with the [O I] 6300\AA\ profile in Fig. 6.   Note that individual emission features closely match between the two profiles.  The particular features to note are the asymmetric emission peak at +680 km s$^{-1}$ in both profiles and the emission peak at -860 km s$^{-1}$ that were absent on the Balmer lines and much less pronounced on the He I profiles.  In the UV, there was a weak, broad emission feature at the  O I 1302\AA\ resonance line.   The persistence of the O$^\circ$ spectrum in novae has been a continuing problem (e.g. Williams 1992) and is considered an effect of shadowing of the central source within the ejecta.  We postpone discussion to the next section.   The [O II] 7319,7330\AA\ doublet was always detected with the same profile as [N II], although severely bended with atmospheric bands and each other.   Although at the end of the STIS spectrum, the O III 3050\AA\ blend was strong during the STIS observation with a similar profile to other permitted lines, especially He I 5876\AA.  The nebular [O III] 4363, 4959,5007\AA\ transitions were present in all spectra, although the first was blended with H$\gamma$.  To obtain a comparison, since the Balmer lines showed a similar profile, we  decomposed the blend at 4340,4363\AA\ by subtracting a scaled H$\beta$ profile (see sec. 4 for discussion).  The result is shown in Figs. 17 and 18 (below, sec. 4.2) for the recovered [O III] 4363\AA\ profile.  The match is almost exact with the other members of the multiplet.   The UV multiplet OII] 1663\AA\ was clearly detected with a similar profile to the N III 4640\AA\  blend.  The line centered at 1402\AA\ appears to be O IV 1401\AA. Neither the Si IV nor S IV resonance lines correspond to the profile centroid wavelength and the interstellar Si IV 1398, 1402\AA\ absorptions confirm that the shortward Si IV emission was absent.   A strong line at 2975.1\AA\ appears to be O IV 2974.42\AA.  The UV recombination line O V 1375\AA\ was not detected.

\begin{figure}
   \centering
   \includegraphics[width=9cm]{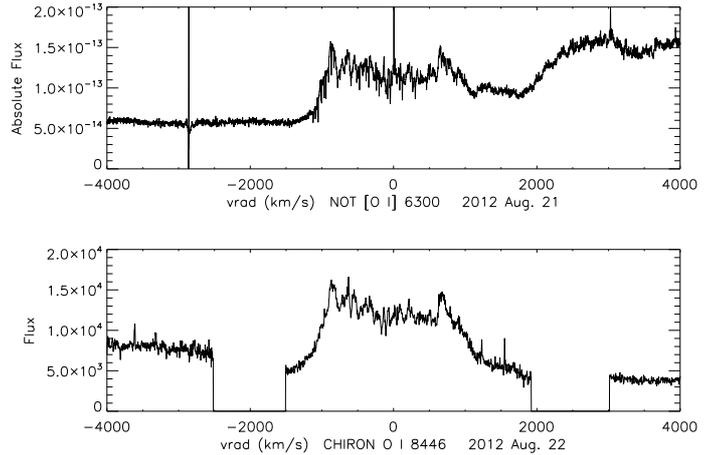}
   \caption{Comparison of the [O I] 6300.3\AA\ (NOT) and O I 8446.3 (CHIRON) line profiles from Nov. 21-22.  See text for details.}
    \end{figure}

\subsection{Fluorine}

An important new result is the identification of the [F III] 2930\AA\ resonance line.  This feature is actually a doublet, 2930.56, 2922.64\AA, that was resolved in late long wavelength high resolution {\it IUE} spectra of RS Oph 1985 and we used those to confirm the identification.  A comparison is shown in Fig. 7;  the narrow emission in the RS Oph spectrum is from the ionized red giant wind (Shore et al. 1996a) and shows the resolved doublet.  The profile in Nova Mon 2012 was the same as O III] 1663\AA\ and other intercombination and forbidden doublets.   There are no resonance lines of this ion in the optical spectrum.   The emission line that is ``missing''  in the {\it IUE}  spectrum is likely Ne V] 2974\AA\ (see also Fig. C.2).  The optical resonance line F II 4789\AA\ was absent at all epochs.  We will return to this in the next section with respect to other ONe novae in the UV.

\begin{figure}
   \centering
   \includegraphics[width=9cm]{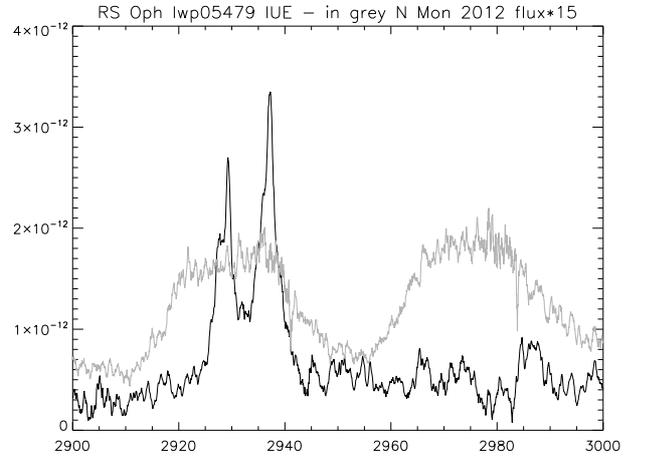}
   \caption{Comparison of the [F III] 2930\AA\ line in the STIS spectrum of Nova Mon 2012 (black)  and the same region in the post-outburst {\it IUE} spectrum  of RS Oph 1985 (gray).  See text for details.}
    \end{figure}

\subsection{Neon, argon}

The main motivation for the STIS observations, the strong similarity of the optical spectra with those of ONe novae, was confirmed by the UV spectra.  Neon is present in three ionization stages between optical and UV transitions: [Ne III] 3869, 3968\AA; [Ne IV] 1602, 2423, 4714, 4725\AA\ and Ne IV] 2974\AA; and [Ne V] 1574; the 1602\AA\ and 1575\AA\ lines are detected  {\it only} in the ONe novae (Fig. 5).   Neither the NOT, SMARTS, nor STIS spectra included Ne III] 3343\AA.  The line profiles showed the same differences as the H I/He II transitions: the higher the ionization stage, the weaker the various narrow emission features in the profile.  The optical [Ne III] line profiles resembled precisely those on the other optical forbidden lines, e.g. [N II] 5755\AA\ and [O III] 5007\AA.    The time development of the optical [Ne III] line is shown in Fig. 8 from the NOT series.  The optical neon lines, especially [Ne V] 3426\AA\ (missing from our spectra), have recently been reported by Munari (2012).

The [Ar III] 7135\AA\ line was probably absent before Oct. 8, after which it was weakly detected (Fig. 8).  The UV resonance line, 3006\AA, was not detected in the Nov. 20 spectrum.   Other optical resonance lines, e.g. [Ar IV] 4711,4740\AA, and [Ar V] 6133\AA,  were searched for but not detected.   There is a feature at 5533\AA\ that is might be the Ar X] 5533\AA\ resonance line.  Its centroid wavelength is not consistent with several other possible resonance line identifications, e.g. Mn VI] 5136\AA, [Cl III] 5137\AA.   This was present in all NOT spectra with the same profile, hence an alternative identification, the excited transition N II 5135,\AA, seems more plausible given the stubborn persistence of the line.   It showed the same profile as O II 6486.46\AA\ on the wing of H$\alpha$.   

  
  \begin{figure*}
   \centering
   \includegraphics[width=16cm]{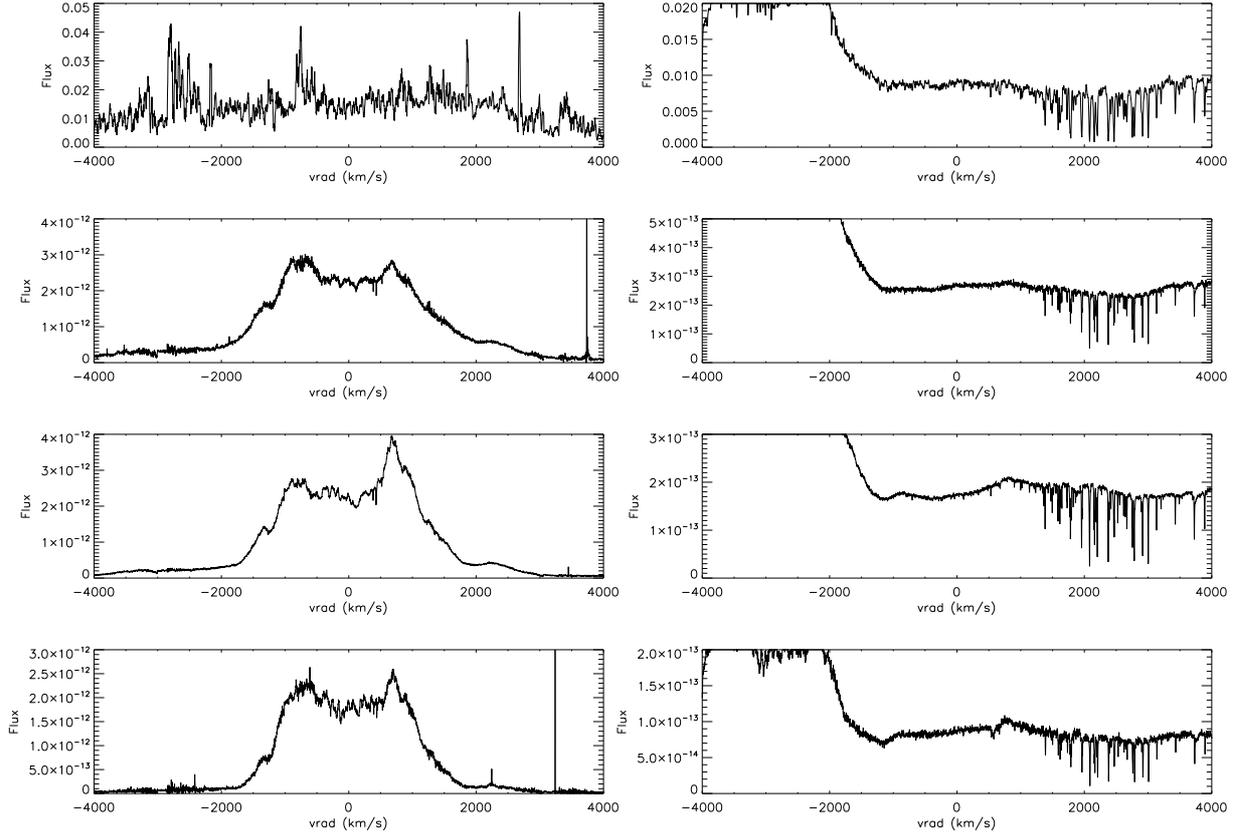}
   \caption{Time series of [Ne III] 3869\AA\ (left) and [Ar III] 7135\AA\ (right) from the NOT spectra; from top: Day 55, 76, 106, and 152.   }
    \end{figure*}

\begin{figure}
\centerline{
\includegraphics[angle=0,width=9.5cm]{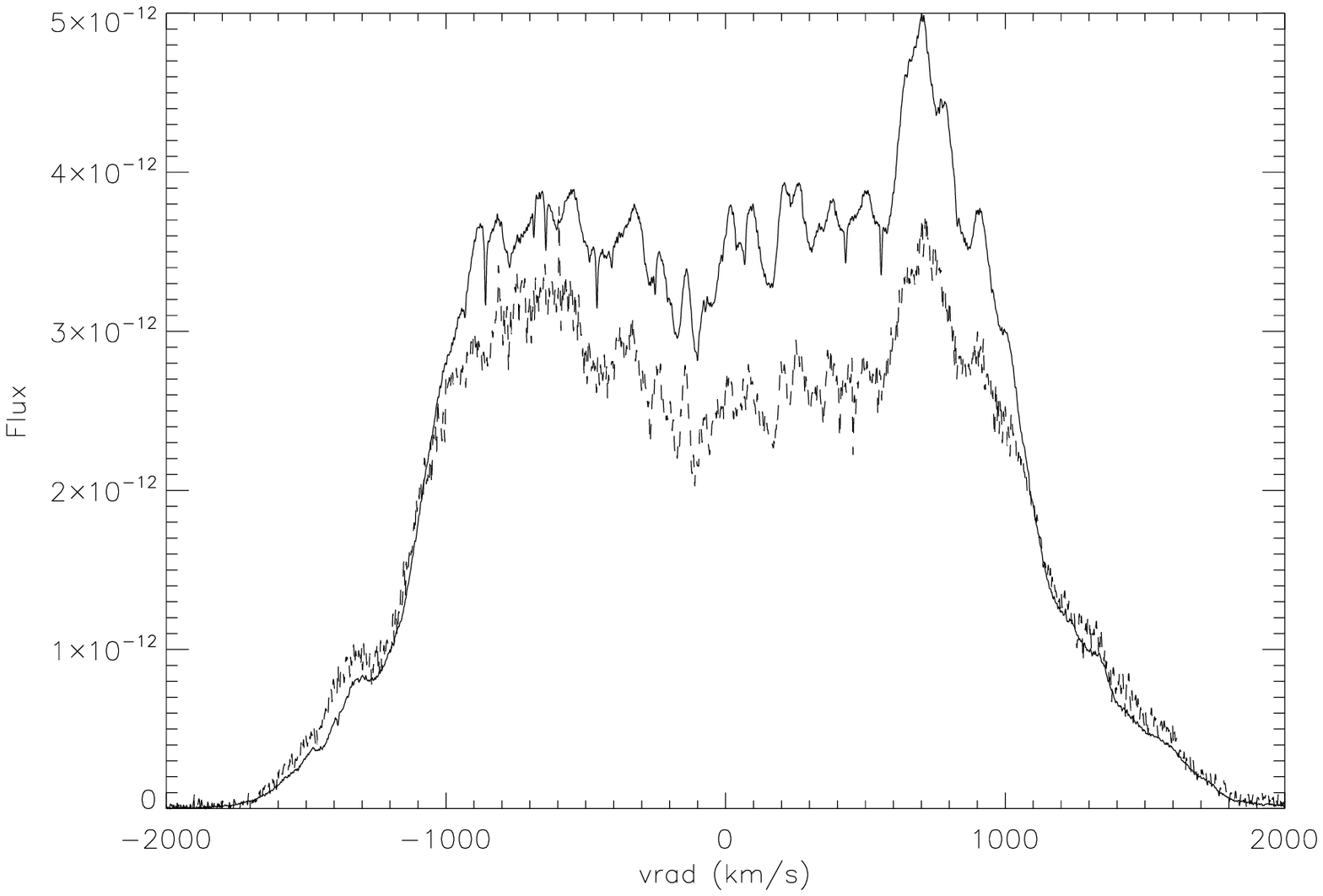}}
           \centerline{\includegraphics[width=9cm]{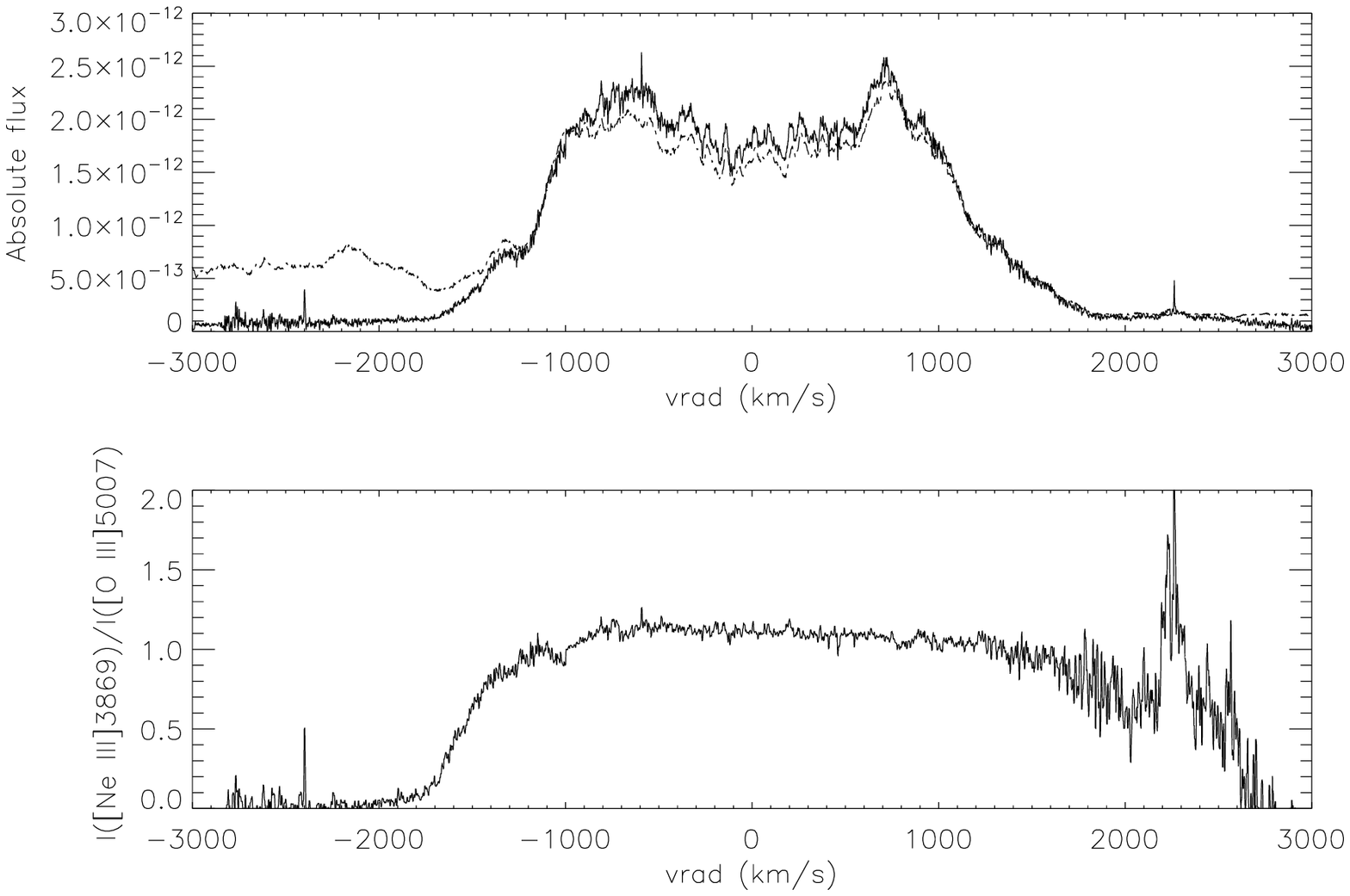}}

\caption{Top: H$\alpha$ (solid) compared to [Ne III] 3869\AA\ (scaled by factor of 1.5) (dash) showing the filamentary structure of the ejecta and possible variation of the Ne/H ratio; Middle: [Ne III] 3869\AA\ and [O III] 5007\AA\ line profiles  compared, bottom: ratio of the neon and oxygen profiles.  This, in cpntrast, shows the likely uniformity of the Ne/O abundances in the ejecta.   See text for discussion. \label{f:many}}
\end{figure}

  \subsection{Heavy elements}
 
The optical Mg I lines were probably detected in the first NOT spectrum on Aug. 16 but faded thereafter.  In contrast, even on Nov. 20 (day 151), the Mg II 2795, 2803 doublet was the strongest feature in the near UV (Fig. C2, appendix C).   The two components were severely blended but structure at high velocity can be discerned on their respective wings that was similar to the high velocity features on the optical lines.  Na I was not detected in emission although, given the late time of the first optical spectra, this is not unexpected.  Interestingly, there was  weak but detectable emission from sulfur, the [S II] doublet at 6716, 6730\AA, that can be recovered by comparison with other weak profiles (see Fig. 10).  The profile, although blended with He I 6678\AA\ on the red, was similar to that for [Ar III] 7135\AA\ and the [Ne III] lines.   The blend on the blue wing of H$\delta$ is consistent with the identification of [S II] 4077\AA\ line but the dissimilarity of the H$\delta$ and other Balmer line profiles prevents the recovery of the full profile as we did for the [O III] 4363\AA\ line.  Regarding the other lines of any sulfur ion, only the [S III] 6312\AA\ line may have been present in the blend with [O I] 6300\AA\ but was very weak and, as previously mentioned, the 1400\AA\ feature is not consistent with a S IV contribution.   This was an siginifantl difference with the most extreme ONe nova, V838 Her, that displayed strong ultraviolet and optical [S II] and S III lines (e.g. Matheson et al. 1993, Schwarz et al. 2007) and a very high S overabundance.  

\begin{figure}
\centerline{
\includegraphics[angle=0,width=8cm]{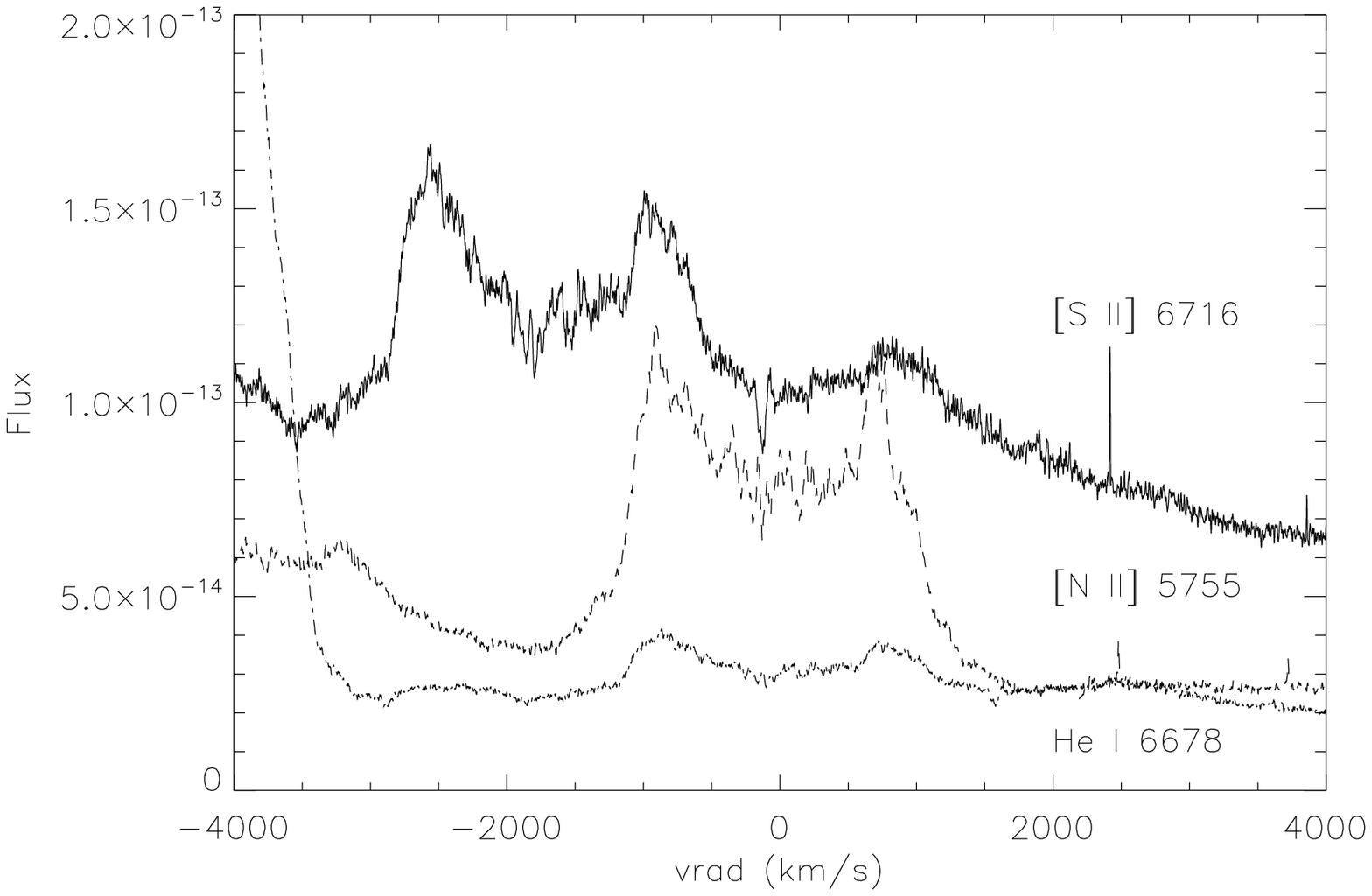}}

            \centerline{\includegraphics[width=8cm]{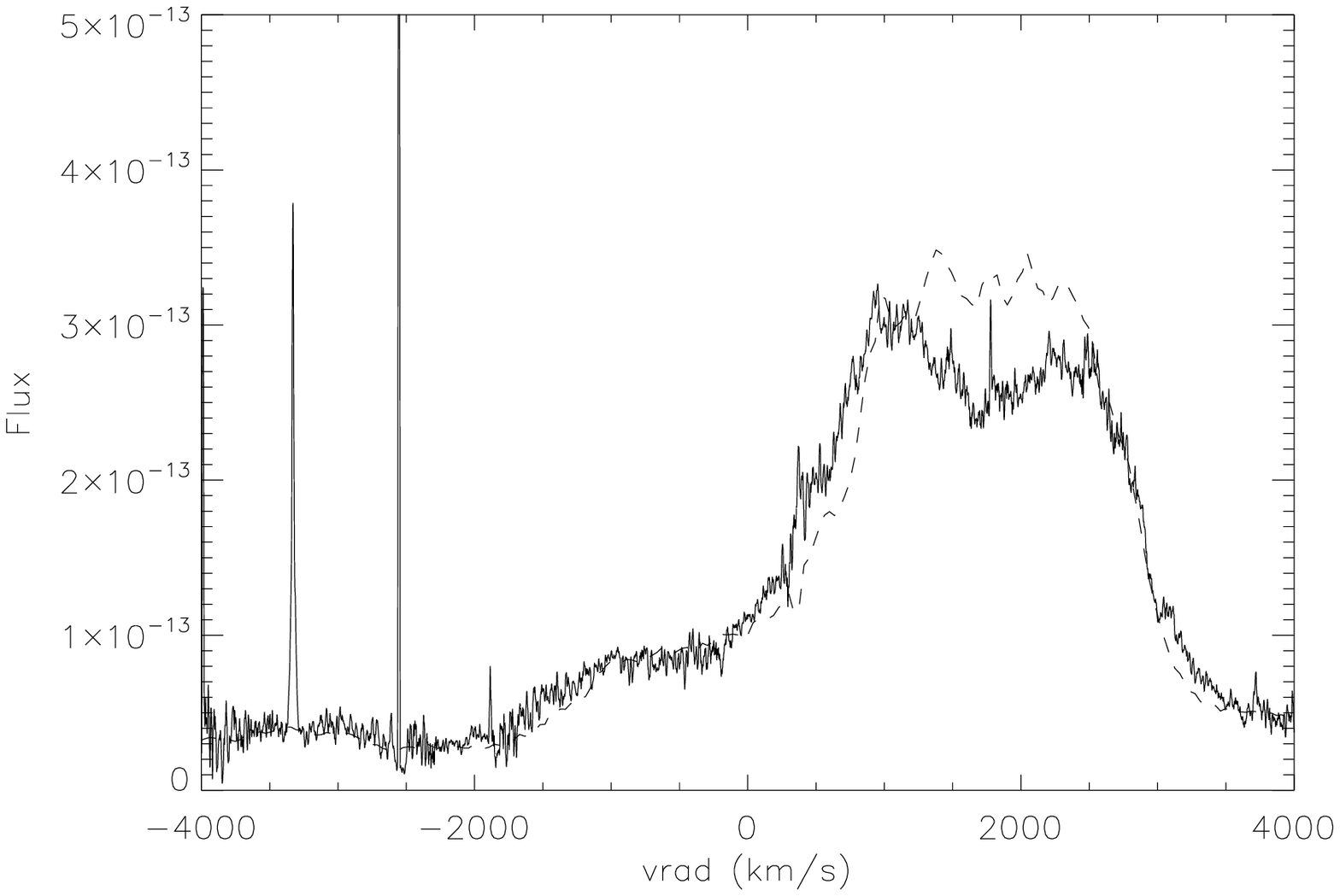}}

\caption{Sulfur lines in the Nov. 21 NOT spectrum.  Top: [S II] 6716, 6730\AA\ lines detected from Nov. 21 compared with the [N II] 5755\AA\ and He II 6678\AA\ lines (both scaled fluxes).  The 6730\AA\ component is in the red wing of [S II] 6716 and the absorption feature is the interstellar diffuse absorption feature at 6616\AA.  Bottom: [S II] 4077\AA\ blended with H$\delta$ (solid) compared with the scaled V1974 Cyg spectrum from 1992 Sep. 25 (dash, see below).   For this comparison the continuum was scaled by a constant, no further corrections were applied. \label{f:many}}
\end{figure}

\subsection{The highest ionization states}

The [Fe VII] 6086\AA\ line was present only in the Nov. spectrum, subsequent to SSS turn-on (see Fig. 11).    The [Fe X] 6374\AA\ line appears to have been present in a complex blend that includes O I 6364\AA\ even before the start of the SSS phase.   The other high ionization iron line, [Fe XIV] 5303\AA, may also have been present in the last NOT spectrum.   

 \begin{figure}
   \centering
   \includegraphics[width=9.3cm,height=10cm]{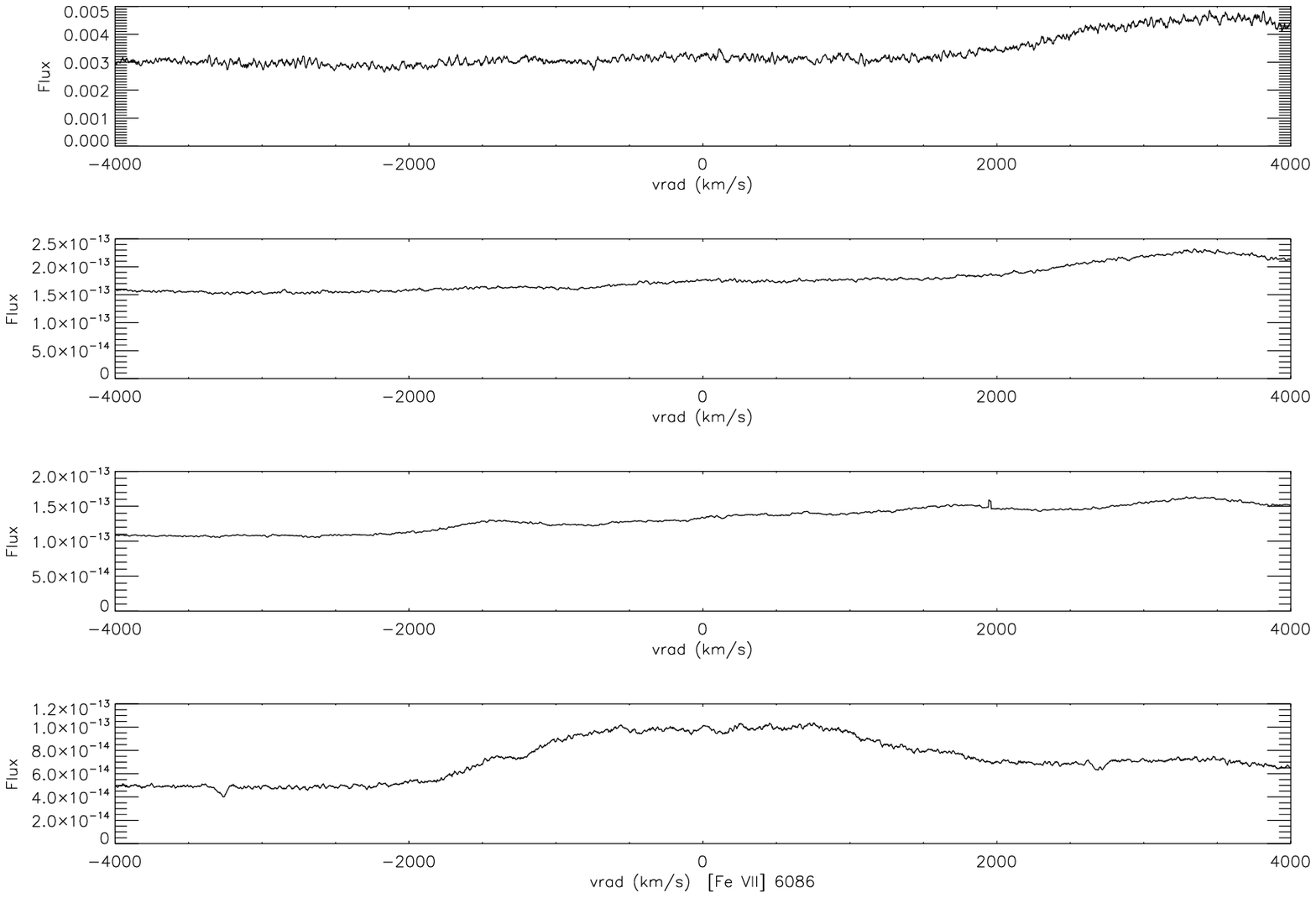}
   \caption{Time series of the [Fe VII] 6086\AA\ line profile from the NOT spectra.}
    \end{figure}        

\section{Determination of extinction and integrated fluxes and comparison with other ONe novae}

Interstellar absorption line measurements toward Nova Mon 2012 were reported by Munari et al. (2012) who, using the E(B-V) vs.  equivalent width calibration for Na I D1, D2  and K I  7699\AA\ ( Munari \& Zwitter 1997) obtained E(B-V)$\approx$0.3.  Based on the comparison of the interstellar Na I D line profiles with the LAB survey 21 cm profile (Kalberla et al. 2005), Cheung et al. (2012b) reported a similar value using the corresponding portion of the H I profile to obtain a hydrogen column density, $N_H$.  However, the UV spectra have been determinative in extending the velocity range of the absorption features, especially indicating that the Na I had a weak component at high (positive) radial velocity (Fig. 12), and the full 21 cm velocity range was spanned by the atomic (and molecular) absorption features.  Thus, using the LAB profile, the line of sight column density is $N_H \approx 5\times 10^{21}$cm$^{-2}$ and E(B-V) = 0.85.  The uncertainty is about 10\% for $N_H$, and about $\pm$0.05 for E(B-V).  We used this to correct the simultaneous  optical and ultraviolet spectra for reddening, the comparison for the two values.  The region between 1800 and 2400\AA\ was very underexposed but the result of the dereddening with standard the value of R=A$_V$/E(B-V)=3.1 is {\it not consistent} with E(B-V)$<$ 0.60.   In our view, the higher extinction is clearly favored even without taking the 2175\AA\ feature into account, and we base all further discussion on the higher value.  For reference, we list in table B1 the equivalent widths of the measurable interstellar lines from the November NOT and STIS spectra.  We show a  sample line profiles in Fig. A.1.

        \begin{figure}
   \centering
   \includegraphics[width=9cm]{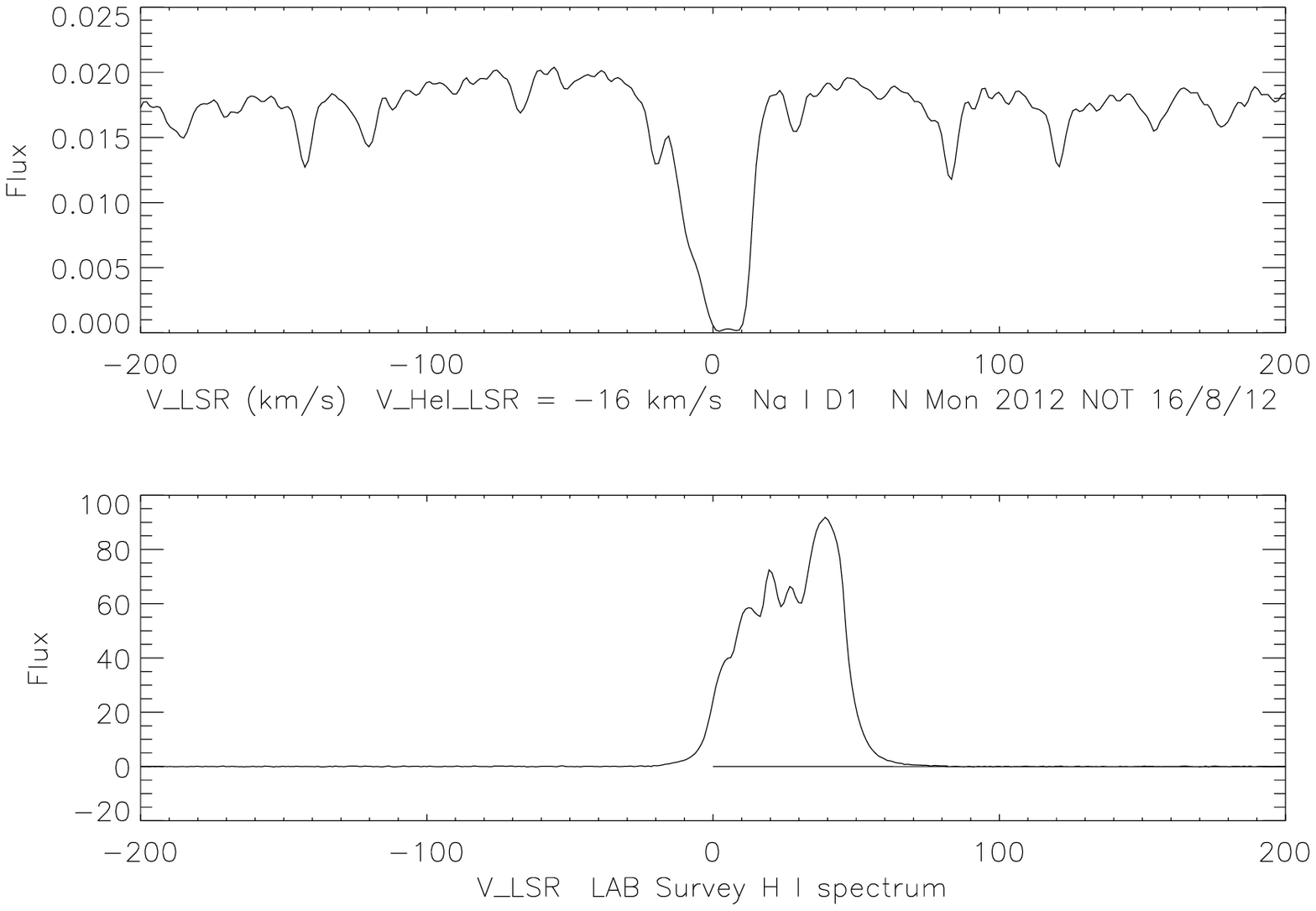}
   \caption{top: Na I 5889, 5895\AA\ from NOT spectra.  Bottom: H I 21 cm profile from the LAB survey.}
    \end{figure}

The immediate result of this correction is important.  The only ONe nova observed in the ultraviolet high resolution at such a late stage was V1974 Cyg.  The spectra were obtained with IUE on day 165, a bit later (relative to peak) than Nova Mon 2012 but after the turn-on of the SSS.  The photoionization re-analysis of this nova by Vanlandingham et al. (2005) find that E(B-V)=0.4 and the best fitting distance to all observations is about 3.6 kpc.  This is slightly higher than the reddening obtained from the early bolometric evolution but consistent (Shore et al. 1994, Shore 2008).   If we deredden the two UV spectra {\it independently}, we obtain the result shown in Fig. 13 along with the comparison of the optical spectra.  The two spectra overlap, {\it without further scaling}.  Thus, it appears that the two are at the same distance with Nova Mon 2012 being substantially more reddened (although this is not particularly surprising given its line of sight).   We note in passing that the {\it WISE} and {\it IRAS} infrared images suggest that Nova Mon 2012 is associated with emission that is likely an intervening translucent molecular cloud. 

{\bf Another consequence of the high reddening estimate here is the resolution of the conundrum posed by Greimel et al. (2012) in their announcement of the detection of the progenitor of Nova Mon 2012 in the IPHAS survey images.  They note that ``If instead an optically thick accretion disk SED is assumed, it would require a reddening of E(B-V)=0.8, the maximum possible for this sightline, and H$\alpha$ EW of 15-20\AA\ to be consistent with the IPHAS colours.''.  We agree with the overall statement, although  at this stage of SSS emission is difficult to distinguish an optically thick disk from that of the continuum emission from the still very hot WD.}

\begin{figure}
\centerline{
\includegraphics[angle=0,width=10cm]{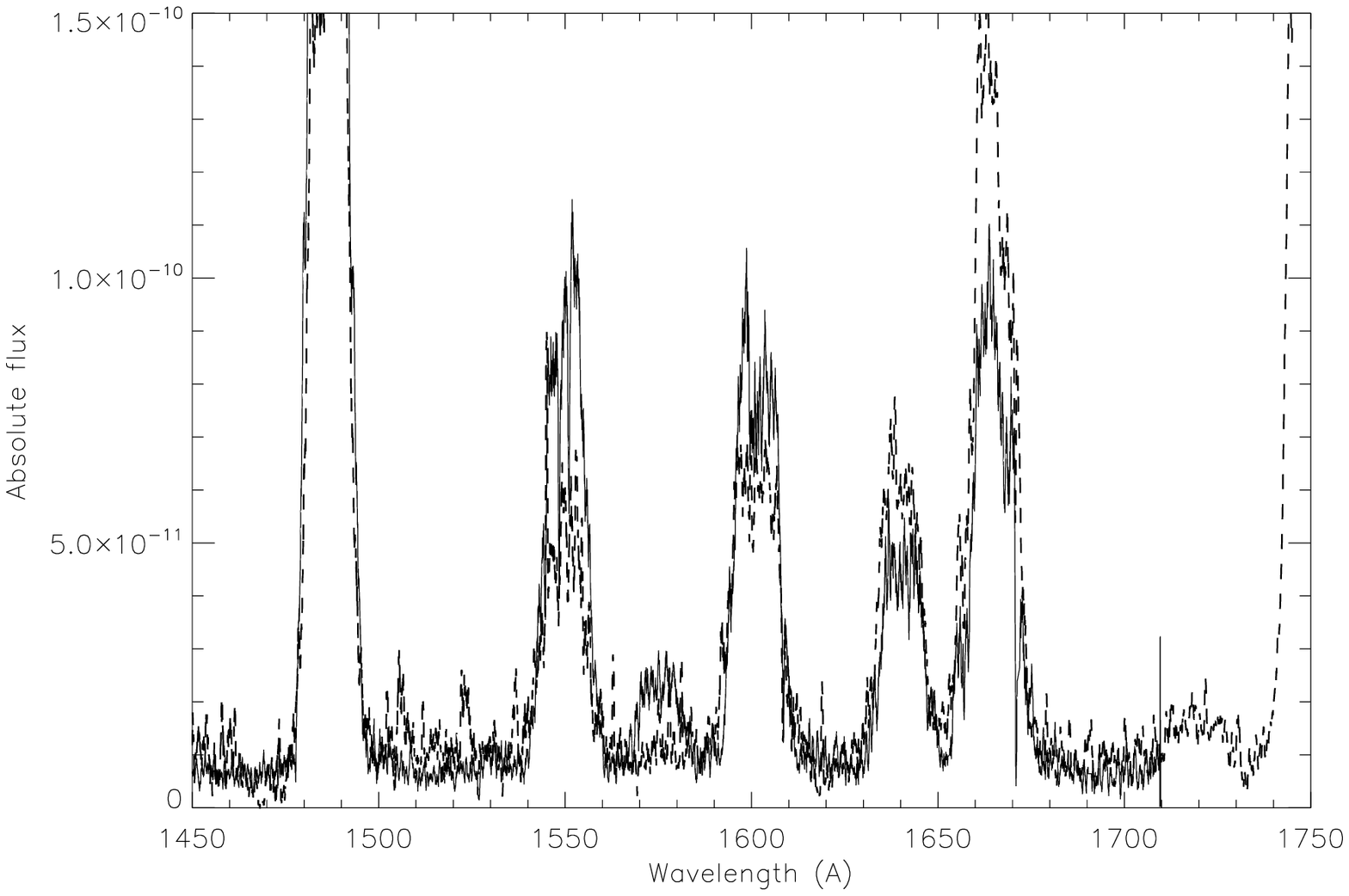}}

            \centerline{
\includegraphics[width=9cm]{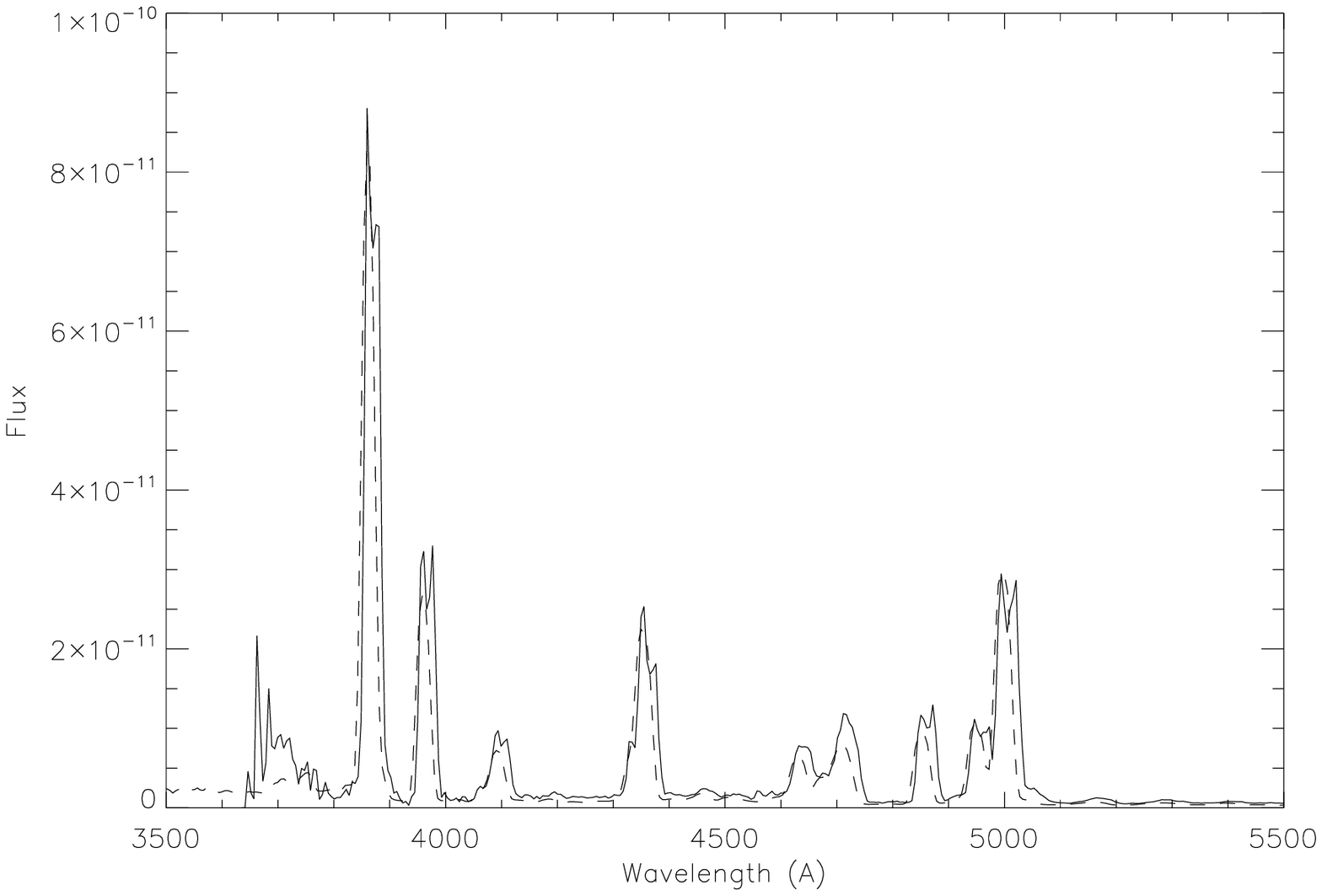}}

\caption{Top: A comparison of Nov. 20 STIS ultraviolet spectrum of Nova Mon 2012, E(B-V)=0.85 (solid) and V1974 Cyg SWP 45245, E(B-V)=0.36, day 165 (dash).  Bottom: Comparison of the optical spectra of Nova Mon 2012 (2012 Nov. 21, NOT; solid) and V1974 Cyg, (1992 Sep. 25; dash), corrected for their respective extinctions.  No scaling corrections have been applied but the NOT spectrum was binned to the resolution of the V1974 Cyg spectrum. \label{f:many}}
\end{figure}

Proceeding to the other two novae for which late time high resolution UV spectra are available -- V382 Vel and Nova LMC 2000 --  we show in Appendix C (Figs. C1 and C2) the galleries of the four ONe novae ordered in the date of the last observation relative to peak.  It is important to note that the individual novae develop {\it spectroscopically} at different rates, this is likely due to differences in the ejecta masses and abundances.  But the continuum emission appears to be the same at the same epoch after outburst.  Significantly, we see that there are transition points in the spectroscopic development, especially in the ultraviolet, that serve to identify both the class and the properties of the ejecta: (1) the disappearance of O I 1302\AA\ and C II 1334,1335\AA; (2) the first appearance of  [Ne V] 1575\AA\ and the strength of the [Ne IV] UV lines; (3) the first appearance in the optical of [Fe VII] 6086\AA.  The first feature corresponds to the stage of complete ionization of the ejecta, the second is due to the high abundance of neon, and the third signals the SSS turn-on.  Several features should be noted in this gallery:
\begin{itemize}
\item The Ly$\alpha$ emission line near 1216\AA\ was far stronger in V382 Vel and Nova LMC 2000 than the other two.  This is explicable assuming a substantially lower interstellar line of sight value of $N_H$ toward these two novae, consistent with the lower extinction.  It cannot be an effect of the observing epoch relative to peak since the ejecta were already sufficiently transparent.
\item The O I 1302\AA\ line was present in the first two and absent in the last two novae.  We do not know precisely when the SSS turned on in V382 Vel but it was already active by the time of the V1974 Cyg observation (Krautter et al. 1996; Shore et al. 1996).  Furthermore, the Nova Mon 2012 observation took place when the SSS was already on, having been first detected several weeks before (Nelson et al. 2012), consistent with the change in the [Fe VII] 6086\AA| line between the Oct. and Nov. NOT spectra).  
\item The [Ne V] 1575\AA\ line was present in Nova LMC 2000, Nova Mon 2012, and V1974 Cyg but not in V382 Vel.  Again this supports a higher ionization stage in the ejecta at the time of the observation.  The rest of the line profiles are quite similar, especially the He II 1640\AA\ line.  The continuum was a bit weaker (also consistent with the light curves) in the  latter two.  
\item The line profiles are quite similar in all four novae.
\item The [F III] 2930\AA\ line was actually detected in the V1974 Cyg long wavelength spectrum, although not noted at the time.
\item Scaling the extinction corrected spectra from 1200 - 2000\AA, since it is only this interval for which there are spectra in common, the continuum  ratio scales as the distance to the source using both V1974 Cyg and Nova LMC 2000 as the comparison.
\end{itemize}

In summary, we propose that the ONe novae form a single unified class, at least at a basic level.  That is, perhaps,  a less ``exploitable'' result than its implication: these novae are from a nearly identical progenitor WD with very similar physical properties for the explosion.

\subsection{Modeling the structure of the ejecta}

The ejecta structure indicated by the inter-comparison of the line profiles was very highly fragmented.  The narrowest confirmed emission peaks (for now called filaments), had FWHM$\approx$10 km s$^{-1}$ in the NOT spectra but the more prominent were a bit broader, 30 km s$^{-1}$. The strongest, broader emission features also appear to have been ensembles of overlapping filaments.  If this is an indication of radial extent, then they individually spanned about 1\% in velocity (hence of the  same order in fractional radius).  There are at least twenty found in the interval from -1500 to 1500 km s$^{-1}$.  These may have changed in their relative intensity between Aug. and Nov. but  most remained discernible from Sept. onward.  The Aug. 16 spectrum was different.  The filaments on several of the prominent nebular lines, and H$\beta$,  had lower contrast but since that spectrum was not absolutely calibrated it is difficult to make more quantitative statements.  These narrow features were roughly symmetric around $v_{rad}=0$ km s$^{-1}$

We used the Monte Carlo procedure described in Shore et al. (2013) to model the line profiles.  Since the nova was observed when all transitions were optically thin, there was no problem with the self-absorption that is an issue at radio wavelengths at the same epochs.   The ejecta were assumed to be either spherical or axisymmetric spheroids (prolate or oblate).  The models have a number of parameters, not all of which are either freely set or  independent.  These are the opening conical (bipolar) angles (an inner angle $\theta_1 \ge$ 0, and an outer angle $\theta_2 \le$ 90$^{\circ}$),  the inner radial extent, $\Delta R/R_{max}$, relative to the outer radius $R_{\max}$ that is given by the maximum velocity $v_{max}$,  an inclination $i$ to the line of sight, and an axial ratio $a$ for the spheroidal and conical models (see Fig 14).  The position angle in the plane of the sky was left as a free parameter but was constrained based on the reported resolved structures from radio interferometry (O'Brien et al. 2012).  A ballistic velocity law was imposed, along with the assumption of constant ejecta mass.  The exponent for the line formation was a free parameter but in the models we discuss here the density was assumed to vary as $\rho(r) \sim r^{-n_{ej}}$ (taking $n_{ej}$=3, constant mass for the ejecta) and a power law density dependence was used for the emissivity, $\rho^{n_d}$.  In general, we assumed $n_d=2$ for  recombination lines.   In the figures, we use the convention of designating the models with  ($n_{ej},n_d,v_{max},\theta_o,\theta_i,\Delta R,i,a$).

 \begin{figure}
   \centering
   \includegraphics[width=9.3cm,height=10cm]{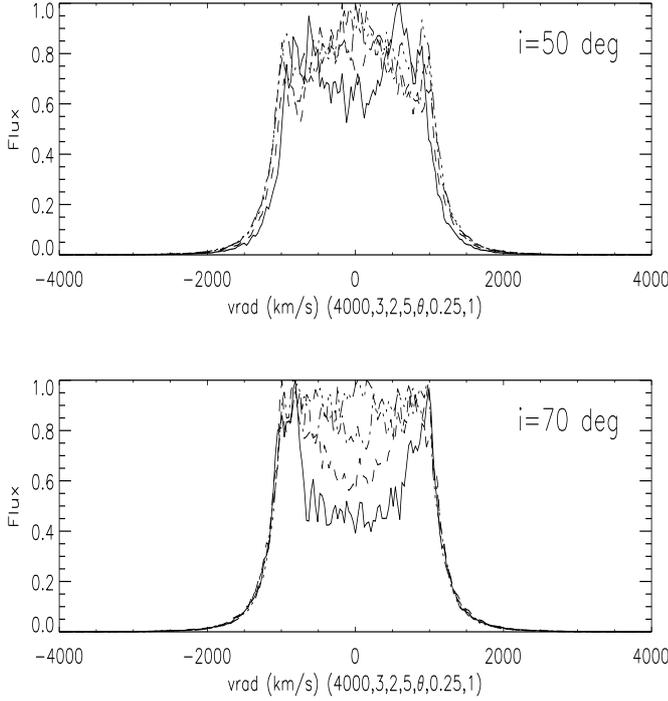}
   \caption{ Examples of high inclination models for bipolar ejecta.  The outer angle was constant, $\theta_o$=5$^\circ$ with the inner angle, here indicated by $\eta$ allowed to vary.  The curves are shown for $\theta$=20 (solid), 40 (dash), 60 (dot-dash), and 90 (double dot-double-dash) degrees for two possible inclinations.   See text for discussion.}
    \end{figure}        

Acceptable fits were obtained for a wide range of inclinations, depending on the assumed values for the cone angles.   Unlike our study of T Pyx, here we have only weaker constraints.  The likely inclination is high, based on the early report of structure from centimeter wavelength interferometry.   As an example of the modeling, we show in Fig. 15 a comparison of the composite set of UV and optical line profiles for a single geometry and two filling assumptions.  The forbidden lines were assumed to form, as in HR Del and other resolved remnants, in a more confined cone than the permitted and recombination lines.  In both cases the maximum velocity, shell relative thickness, and inclination were kept fixed as was the outer cone angle.  The He II 1640\AA, for instance, appears to be formed in a larger volume of the ejecta (Fig. 16) than, for example, the [N II] 5755\AA\ line.  

\begin{figure*}
   \centering
   \includegraphics[width=17cm]{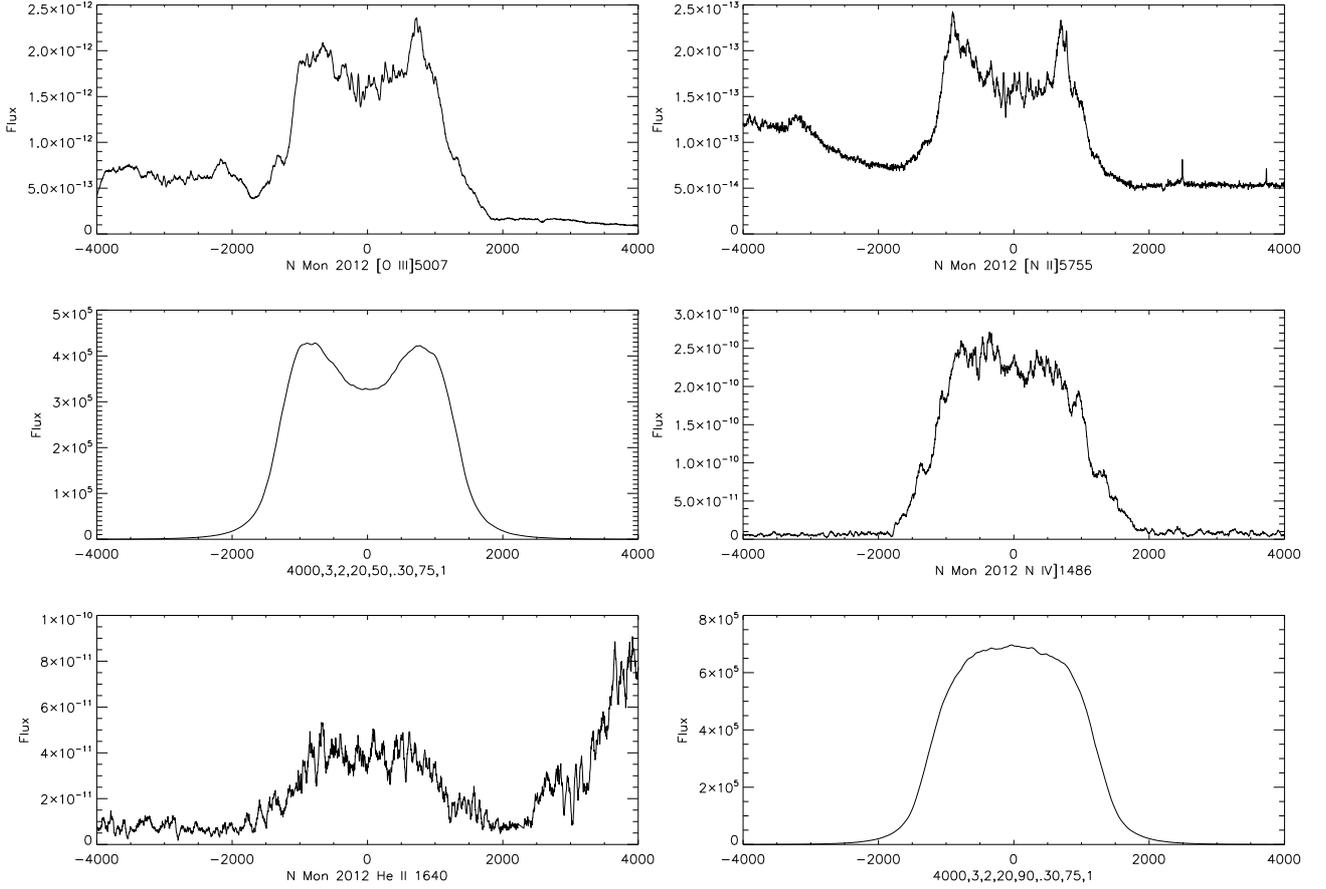}
   \caption{Comparison of model profiles with the observations from Nov. 20-21.  These model comparisons show an example of the range of possible fits.}
    \end{figure*}

A further indication of the ejecta structure is provided by a comparison of the profiles for lines of different ionization stages. of the same element  This is most easily effected using neon.  There is a systematic change in the profiles moving from the lowest ionization, {[Ne III], to highest, [Ne V].  The change can be reproduced using a progressively greater inner angle, varying as shown in Fig. 14.  The same holds for He I and He II, and for [N II] to N IV].    It should be noted that in late epoch infrared observations of V382 Vel with {\it Spitzer}, Helton et al. (2012) found the same behavior for the infrared neon lines (see their Fig.   5).  This cannot be modeled with standard photoionization codes, such as {\it Cloudy}, since they generally assume spherical symmetry.  Monte Carlo codes such as MOCASSIN (e.g. Ercole et al. 2003) are required but can be constrained by the line profile modeling.

\begin{figure}
\centerline{\includegraphics[angle=0,width=4.5cm]{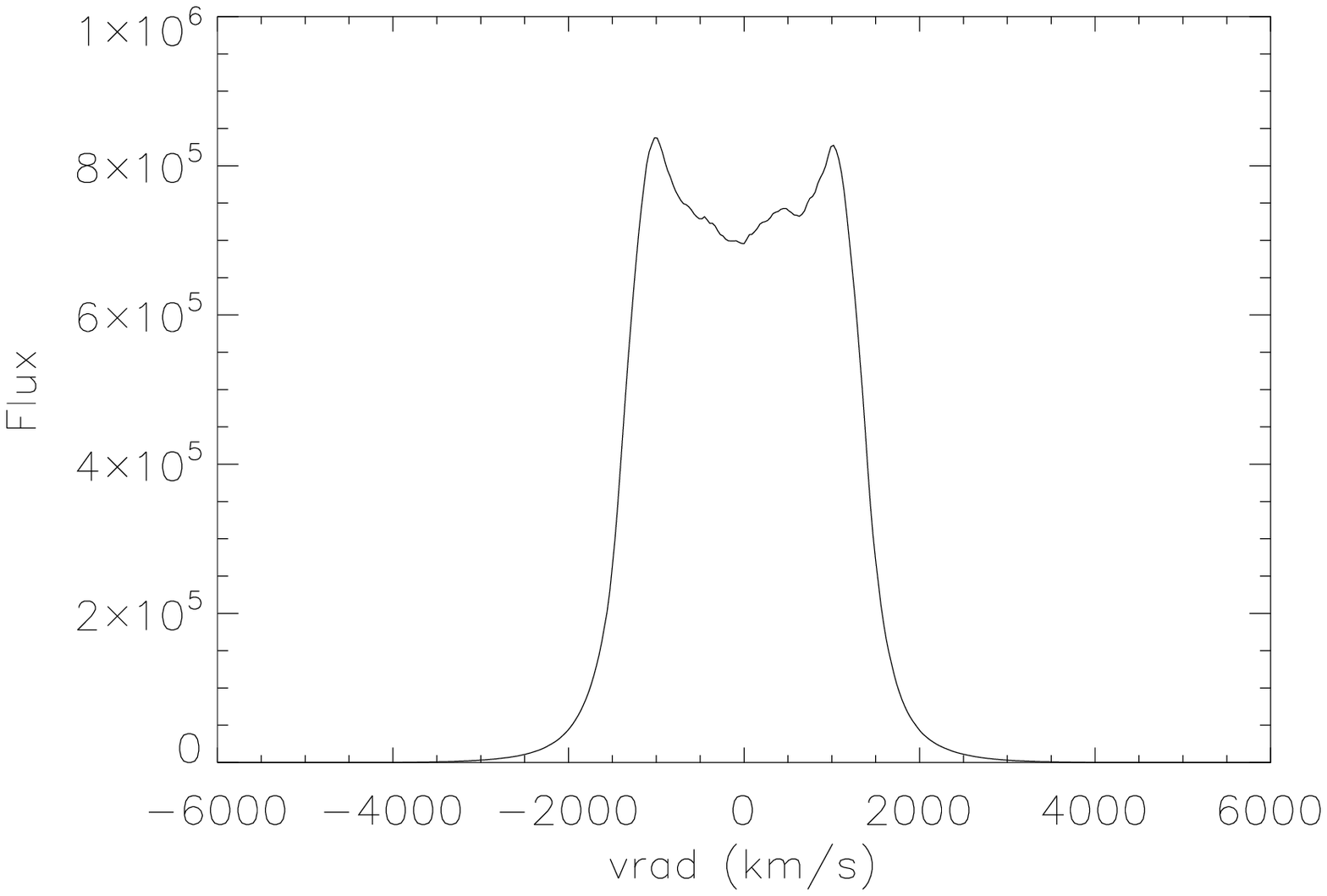} \qquad
            \includegraphics[origin=ltangle=0,width=4.5cm]{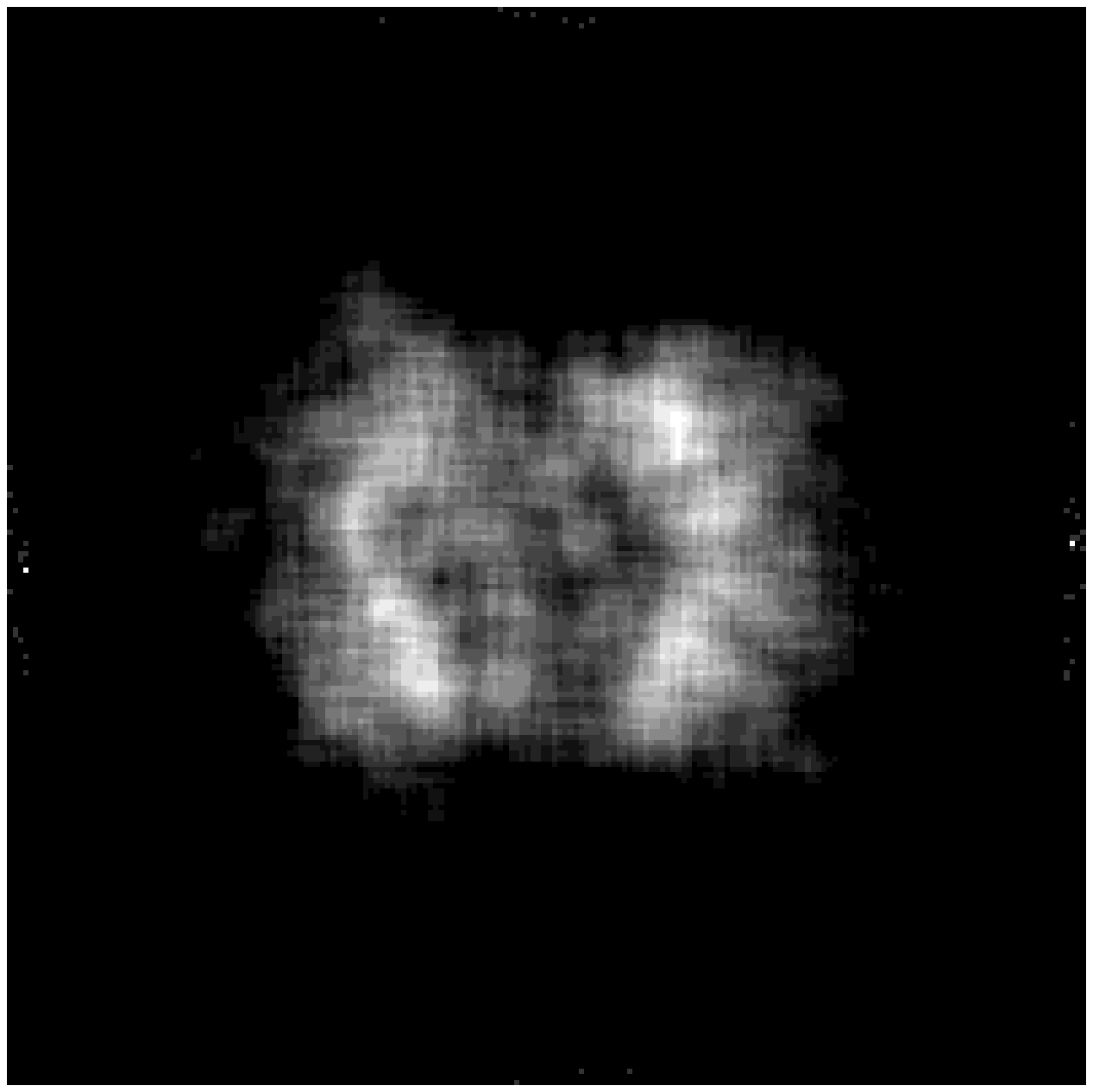}}  
\caption{Left:  Model spectrum for Balmer and permitted lines for Nova Mon 2012 for v$_{rad}=$4000 km s$^{-1}$, $\theta_o=$5$^\circ$, $\theta_i$=60$^\circ$, $\Delta R/R=$0.3, $i$=70$^\circ$.  Right: pseudo-image of the ejecta corresponding to the parameters used for the profile simulation (the image is cut at 30\% of the peak flux). \label{f:many}}
\end{figure}
\subsubsection{Luminosity}

The integrated flux between 1200 and 7400\AA\ dereddened with E(B-V)=0.85  was $2.5\times 10^{-8}$ erg s$^{-1}$cm$^{-2}$.  Thus, the {\it lower limit} for the bolometric luminosity on Nov. 21 was  $> 9.8\times 10^3$L$_\odot$ for a distance of 3.6 kpc.   This implies that, if the WD is at the Chandrasekhar mass and the total luminosity is less than the Eddington luminosity, the EUV and XR accounted for  $\leq$2.3 times the measured emission from the 2012 Nov.  observations.

\subsection{Electron density and mass of the ejecta}

For every epoch, the [O III] 4363, 4959, 5007 \AA\ lines were strong. The [O III] 4363\AA\ line was, however, severely entangled with H$\gamma$.  Using a similar procedure that was used for the TPyx analysis outlined in Shore et al. (2012), we assumed the similarity of the Balmer profiles and scaled H$\beta$ to the dereddened blend, subtracted it, and shifted the residual profile to the [O III] 4636\AA\ line rest wavelength.    We thus obtained the velocity dependent (projected line of sight) electron density from the ratio of the profiles (Figs. 17 and 18).  The derived density depends on the electron temperature, T$_e$ (Osterbrock \& Ferland 2006).   Unfortunately,  for Nova Mon 2012 we have no other simultaneous diagnostic, as we did in T Pyx using the [N II] isoelectronic lines, to  guide the choice of the temperature so we used $T_e = 10^4$K.  This yields  $n_e \approx 3\times 10^7$ cm$^{-3}$ for the innermost portion of the ejecta for the Nov. NOT spectrum.  We note that the ratio, and therefore the value of $n_e$ among the individual emission peaks, varies by less than a factor of three.

The measured integrated H$\beta$ flux on Nov. 21 was 2.3$\times 10^{-11}$ erg s$^{-1}$cm$^{-2}$, corrected for extinction this was 4.0$\times 10^{-10}$ in the same units.  The H$\beta$ luminosity was 8$\times 10^{36}$ erg s$^{-1}$ for a distance of 3.6 kpc.  Since the ejecta electron density was $n_e \approx 3\times 10^7$ cm$^{-3}$ at an inner radius of 1.7$\times 10^{15}$cm, taking T$_e \approx 10^4$K gives a required volume of  $\approx 8\times 10^{45}$cm$^{-3}$; hence, the filling factor is $f \approx 0.3$ using standard Balmer line recombination coefficients (P\'equinoit et al. 1991; Osterbrock \& Ferland 2006).  This is somewhat higher than the filling factor adopted for V1974 Cyg in Vanlandingham et al. (2005).   It is likely  an upper limit since the filamentation evinces an even lower value for $f$.  Therefore, the mass of the ejecta  is  $M_{ej} \le 2\times 10^{-4}f$M$_\odot$.    An independent estimate using the {\it Cloudy} models gives a similar result and the mass is the same order as derived for V1974 Cyg and V382 Vel.  The main differences between this and previous estimates are the {\it  independent} derivation of the electron density profile in the ejecta and the late stage of the observation.  It should be noted that because the SSS was active, the recombination was not in the expansion-dominated stage (see Shore et al. 1996, Vanlandingham et al. 2001) so standard photoionization analysis is still valid.

  \begin{figure}
   \centering
   \includegraphics[width=9cm]{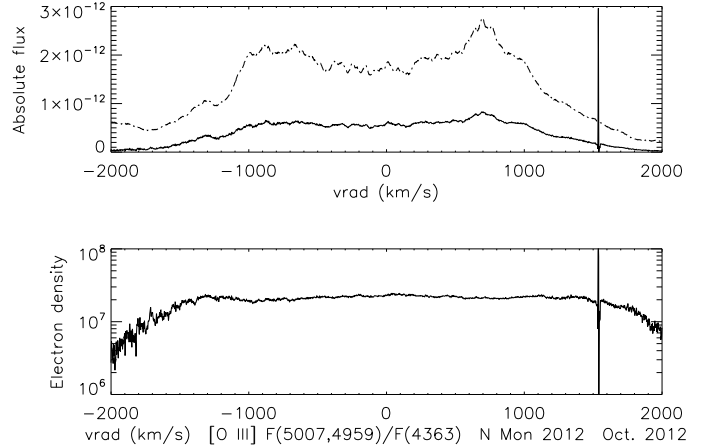}
   \caption{ Electron density [O III] diagnostic analysis for Oct. 2012 based on NOT spectra using the [O III] line ratios.  Fluxes are corrected for E(B-V)=0.85 (see discussion).   Top: [O III] line profiles: 4363\AA\ (solid), disentangled from the blend with H$\gamma$, 4959+5007\AA\ (dash).  Bottom: derived $n_e$ as a function of radial velocity.  For $|v_{\rm rad}|>2000$ km s$^{-1}$ the profiles are to weak for the ratio to be meaningful.  }
    \end{figure}

\begin{figure}
   \centering
   \includegraphics[width=9cm]{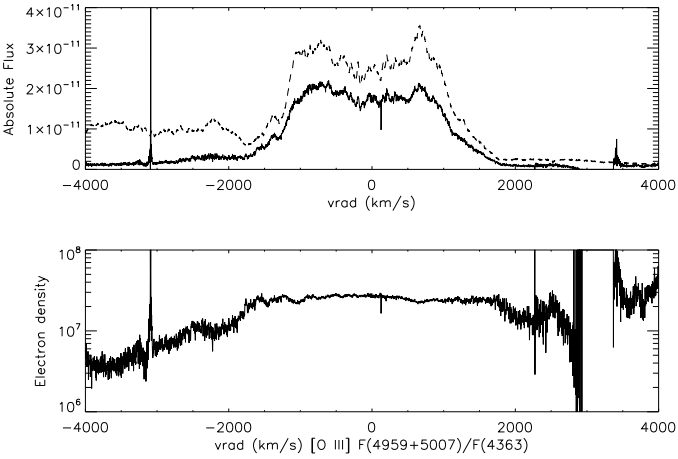}
   \caption{Same as Fig. 17 for Nov. 2012,  based on NOT spectra.   Top: [O III] line profiles: 4363\AA\ (solid), disentangled from the blend with H$\gamma$, 4959+5007\AA\ (dash).  Bottom: derived $n_e$ as a function of radial velocity.  }
    \end{figure}

\subsection{{\it Cloudy} photoionization analysis}
\begin{figure*}
   \centering
   \includegraphics[width=17cm,height=10cm]{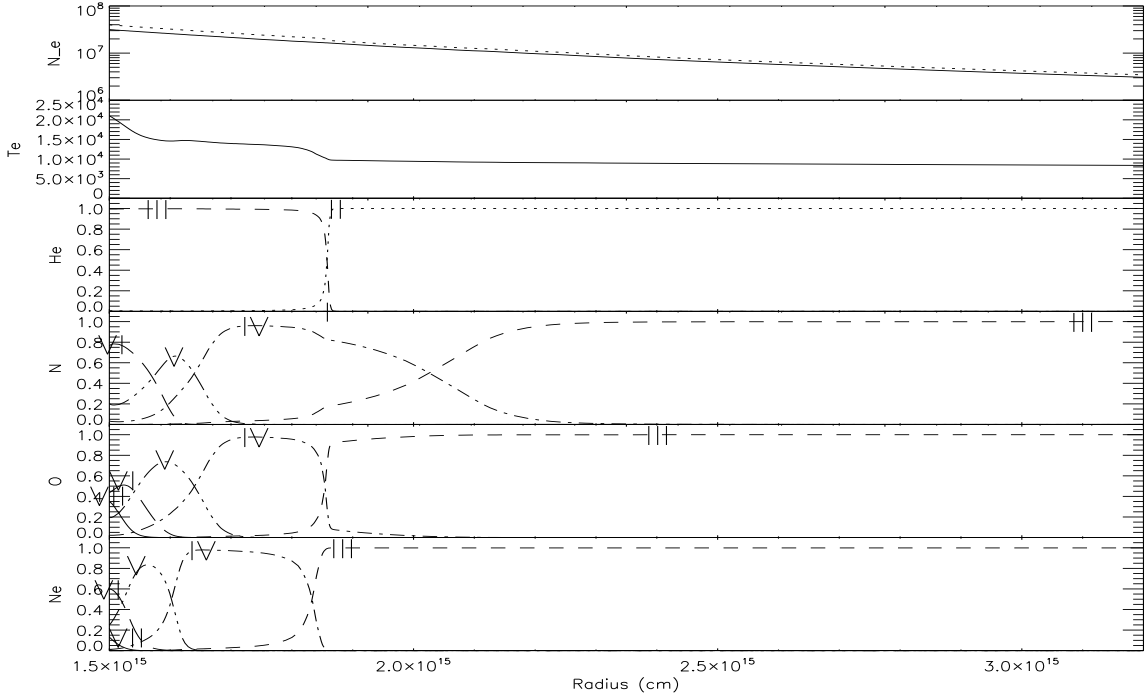}
   \caption{Resulting ionization structure for the {\it Cloudy}  photoionization model  with parameters $\log L = 38.0$, $n_e=3.1\times 10^7$ cm$^{-3}$, $R_{min}=1.5\times 10^{15}$ cm, $R_{out}=6\times 10^{15}$ cm, ({\it covering factor}) = 0.8, $f$=0.3, T$_{\rm eff}=3\times10^5$K.  The abundance set is the same as that used for V1974 Cyg.   Note, in particular, the ionization distribution for Ne and N.   See text for discussion.}
       \end{figure*}    
       
As a test of the hypothesis that we are dealing with a homogeneous subclass, we used the abundances derived in previously published analyses  for V1974 Cyg (Vanlandingham et al. 2005) and V382 Vel, (Shore et al. 2003) to predict the integrated line fluxes for the ejecta parameters we determined for Nova Mon 2012.  The reader should note that these are {\it not} attempts to obtain a photoionization model that fits the observations.  It is, instead, a way of contrasting the forward-predicted expectations with the data.  In this way we can identify the most likely abundance differences in a quantitative manner.  The abundance set (logarithm of the number relative to H) was He=-1.00, C=-3.61, N=-2.52, O=-2.34, F=-6.54, Ne=-2.32, Na=-5.69, Mg=-3.70, Al=-4.20,  Si=-4.75, S=-4.79, Ar=-5.40.  The heavier elements were fixed at solar values.   These are similar to abundance results for several other novae, including V351 Pup (Saizar et al. 1996) and Nova LMC 1990 Nr. 1 (Vanlandingham et al. 1999).   The source luminosity and effective temperature were estimated from reports of the {\it Swift} fluxes and spectra (e.g. Nelson et al. ATel 4590) and our estimate of the luminosity. The lines measured on Nov. 21 are tabulated in table 3a of appendix B along with the predicted fluxes.   We also provide the time variations for the measured integrated line fluxes of lines in common with Munari (2012) from the NOT spectra in table 3b; the flux uncertainties, about 10\%, were dominated by the calibration.  The model ionization stratification is shown in Fig. 19. The parameters derived from the profile modeling were: the inner boundary electron density, n$_e(R_{in})$ =  3.2$\times 10^7$ cm$^{-3}$, the inner radius $\log R_{in}$ = 15.18, $\Delta R/R_{out}=0.5$, a  cover factor of 0.8, and a filling factor ($f$) of 0.3.  The source temperature (assumed for a Planck distribution) T$_{eff}$ = 300,000 K for a luminosity of 2.5$\times 10^4$L$_\odot$.    This model  gives $\chi^2$/dof = 68.5/29, but most of the variance comes from a few poorly fit lines, e.g. the Mg II 2800\AA\ doublet that is probably formed in regions shielded from the central source (Williams 1992).  The reduced $\chi^2$ was substantially larger for a higher source T$_{\rm eff}$ (e.g. for a 400 kK model, it increased by a factor of 5) but this is used here to quantify the agreement and {\it not} as a goodness-of-fit statistic for a full analysis that will be presented in the next paper.

\subsection{Comparisons with other ONe novae: spatial structures and line profiles}

Our distance, extinction, and luminosity determined for Nova Mon 2012 are in accord with the reported early radio interferometric structure determinations.  Based on our modeling of the profiles and the time of maximum, we would expect that the radio would have shown a separation between the components of about 40 mas at around 80 days after outburst. This assumes v$_{\rm rad,max} = 4000$ km s$^{-1}$ and that the main contribution to the emission from the inner radius, $\Delta R/R \approx 0.3$.    The reported separation at cm wavelengths was 40 mas (O'Brien et al. 2012).   The detection of separate structures also suggests a high inclination, although the precise value has not yet been reported.  As a prediction, we include the synthetic image that is consistent with our Balmer line models in Fig. 16.

\begin{figure}
\centerline{
\includegraphics[angle=0,width=8cm]{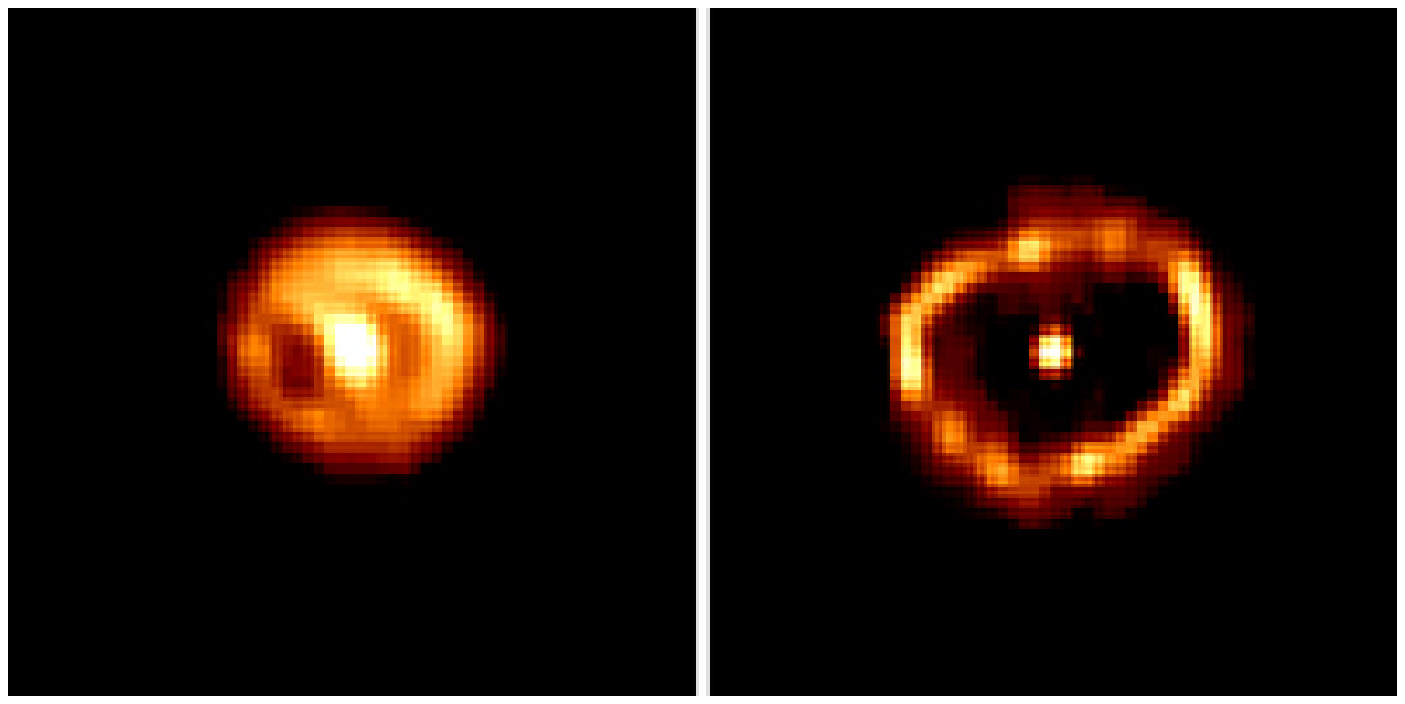}}
            \centerline{\includegraphics[width=4.0cm,angle=0]{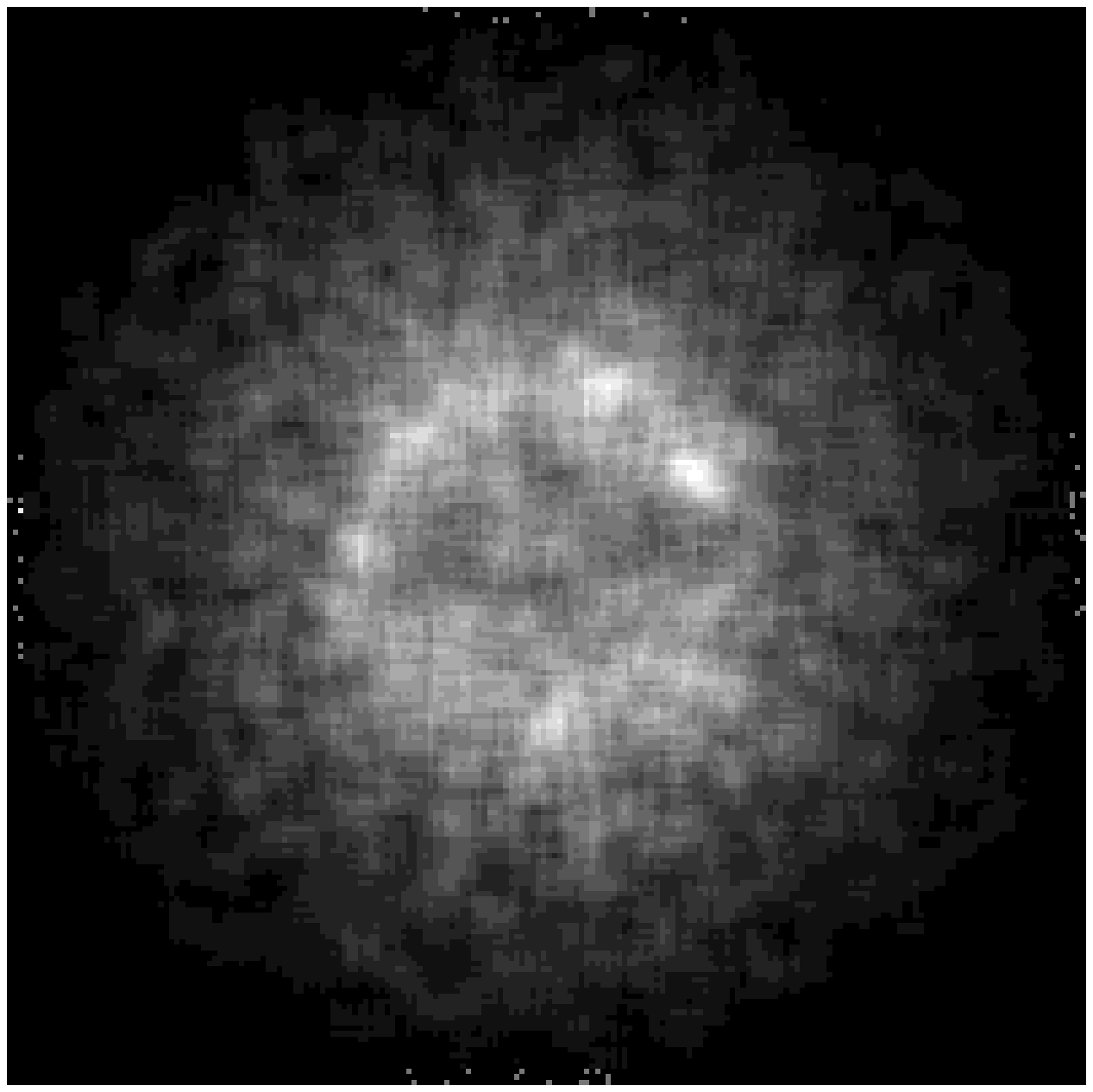} 
            \includegraphics[width=4cm,angle=0]{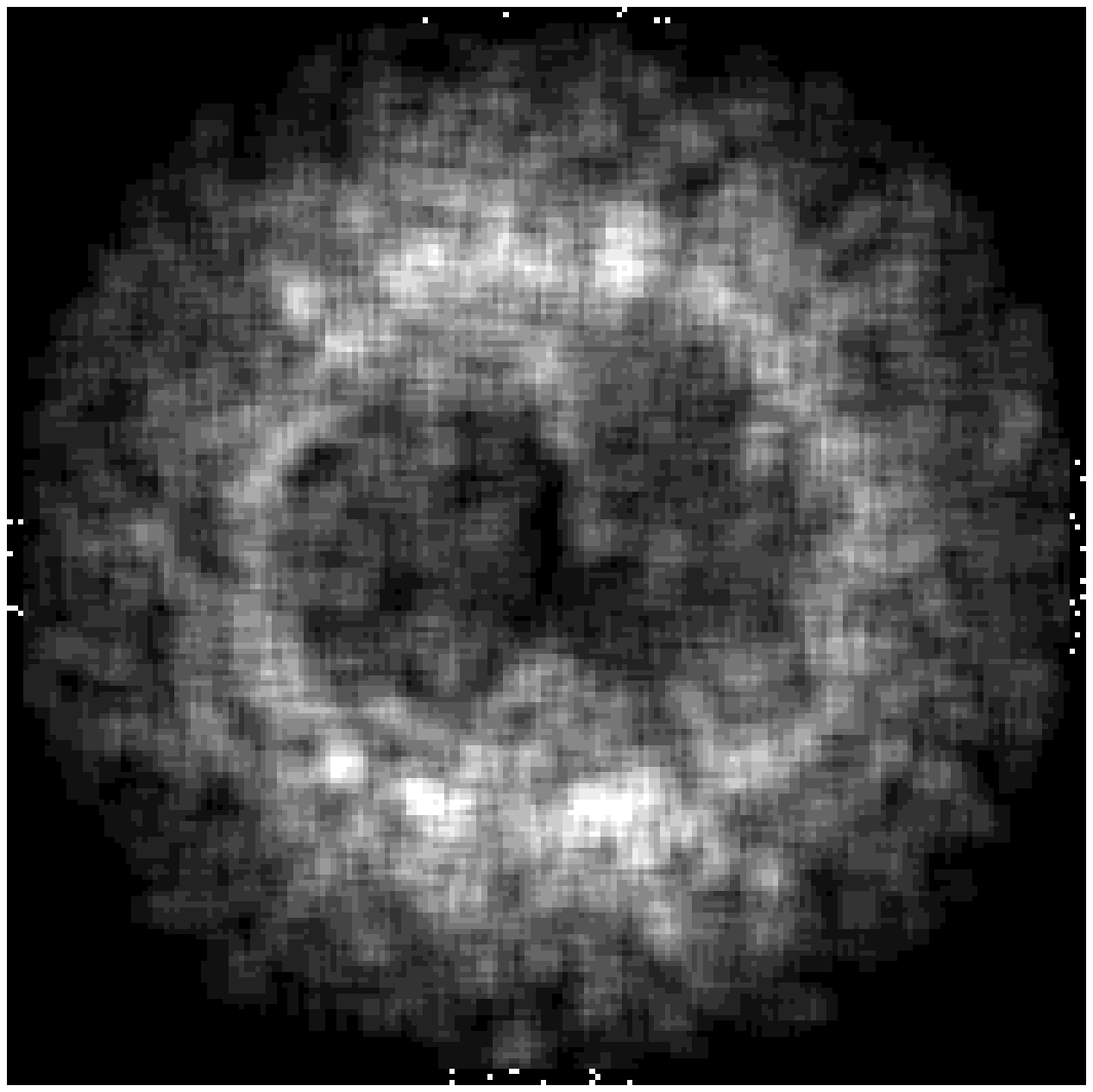}}
\caption{Top:  {\it HST}/FOC images of the ejecta of V1974 Cyg from days 464 (left, F274M) and 723 (right, F501M).  The first image was before installation of COSTAR.  Bottom: sample model ejecta -- left, first epoch with parameters  (3,2,10,80,0.3,35,1); right, second (epoch with (3,2,10,60,0.5,35,1) . \label{f:many}}
\end{figure}

As a check on the procedure, we compare the predictions (or post-dictions) of the models for V1974 Cyg with the observations of the resolved ejecta from the HST archives (Paresce et al. 1995, Chochol et al. 1997).  These are shown in Fig. 20 and a sample of optical line profiles compared with the models is shown in Figs. 21 and 22.  The first observation, obtained about 464 days after outburst, showed a bar-like structure that was unexplained at the time (1992 Sept.) but presumed to be an artifact.  Instead, the resolved narrow ring from around day 723 showed several, probably unresolved, emission knots and an oval shape.   Both are consistent with the expected expansion.  If the inclination is about 40 $^\circ$, the bar-like feature in the first image is explicable as a projection effect of the lobate structures for an inner angle of about 80$^\circ$.  The second image is consistent with this inclination provided the emission is coming from the innermost part of the ejecta, with $\Delta R/R \approx 0.2$.  Again, this is consistent with the expansion rates.  {\bf This is very difficult to understand for a planar configuration but the solution for the emission lines, axisymmetric optically thin ejecta with an inclination $i$ from 30 to 40 degrees to the line of sight, naturally explains both observations.  The measured expansion was not linear in time as would be expected from a naive application of a ballistic velocity law.  However, because of the interval between the two epochs and the decreased emissivity of the outer ejecta, the innermost region shifted toward the central star resulting in an decrease in the observed shell radius of about 40\% relative to constant velocity.    Small aperture {\it GHRS}  observations of one of those knots showed that the Ne/C ratio was different than the integrated large aperture spectrum and two velocity components for He II 1640\AA\  resolved within this single knot (Shore et al. 1997) and separated by about 1000 km s$^{-1}$.  This is inexplicable assuming the original ring geometries proposed for the ejecta.  It is a natural consequence, however, of the superposition of the oppositely directed bipolar lobes along the line of sight (see Shore et al. (2013) for further discussion). }

\begin{figure}
   \centering
   \includegraphics[width=9cm]{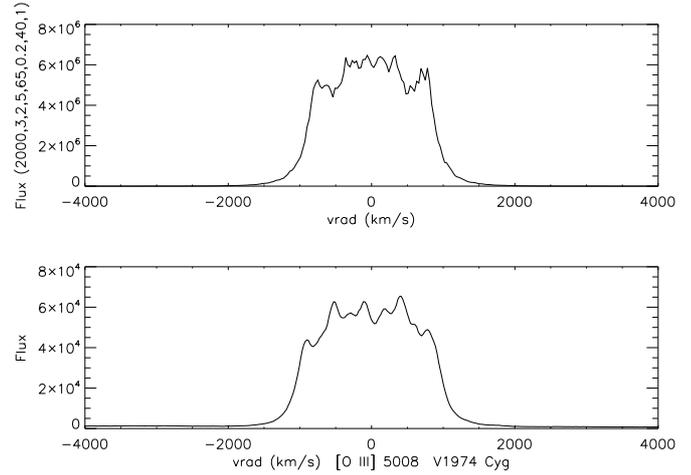}
   \caption{Comparison of model line profile with the H$\alpha$ profile from 1992 Sep. 25 (medium resolution, Perkins 1.8m Lowell Observatory spectrum courtesy of R. M. Wagner, see Austin et al. (1996) for observational details.  The model parameters are indicated. }
    \end{figure}

\begin{figure}
   \centering
   \includegraphics[width=9cm]{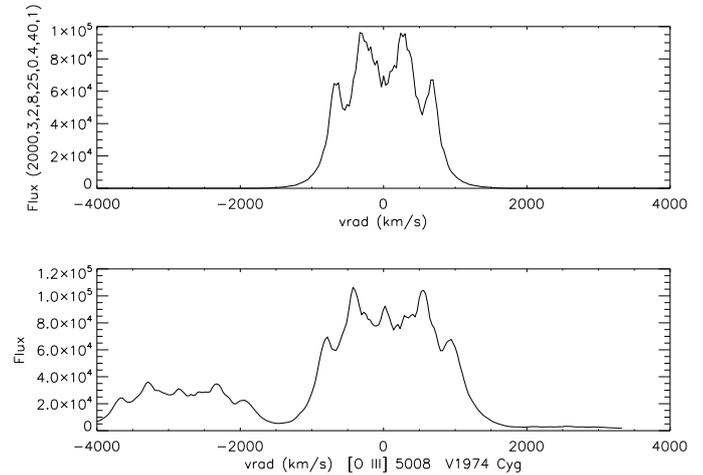}
   \caption{Comparison of model line profile with the V1974 Cyg [O III] 5007\AA\ profile from 1992 Sep. 25 for parameters indicated. }
    \end{figure}
    
The ejecta of V382 Vel have not yet been resolved but they should be visible.  The mass was similar to that we derive for Nova Mon 2012 and that was  previously  found for V1974 Cyg (Shore et al. 1993, 2003).  The late epoch line formation was dominated by the effects of the expansion on the recombination rate, hence the ionization and line emission should freeze out within several years after the ejection.   The likely smaller distance to this nova, and the similarity of its line profiles to the other ONe novae, suggest that it, like V1974 Cyg, may have an intermediate inclination.   We further note that one other ONe nova has spatially resolved ejecta, QU Vul (Downes \& Duerbeck 2000; Gehrz et al. 2008).  In this case, there is the  additional constraint that the binary system is eclipsing.  Thus the combination of line profiles and imaging are consistent with a high inclination.    We will present a more extensive analysis of the combined line profiles and ejecta structures in a future paper.

\section{Discussion: Nova Mon 2012 as a $\gamma$-ray source}

We now come to the core result of this study: {\it the four ONe novae for which high resolution panchromatic observations are available form what appears to be a single class}.  Despite the lack of previous observations of the HE $\gamma$-ray emission from the other three, indeed also for other classical novae (with the exception of Nova Sco 2012, see Cheung et al. 2012), we conjecture that the emission is characteristic of the early stage of the outburst.  It is well known that novae in their early, optically thick stages are hard X-ray  sources and only after the ejecta are transparent do they appear as supersoft X-ray emitters.   A simple explanation for the X-rays is internal shocks caused by the collision of the filaments that eventually freeze out in the expansion.  In the earliest stages, this is a natural source for particle acceleration, the differential velocities are of the same order as the velocity width of the ejecta, i.e. several thousand km s$^{-1}$.   A complex ejecta structure and distribution of both abundances and ejection velocities and times is indicated by highly structured, turbulent three dimensional modeling of the start of the explosion (Casanova et al. 2011).  While the available multidimensional models are still too few and specialized to generalize, two dimensional models have recently been proposed for ONe explosions (Glasner et al. 2012) that confirm the extension to the more energetic explosions, albeit with many physical and computational limitations.   {\it The detailed mechanism is fundamentally distinct from that responsible for the HE emission observed from V407 Cyg}  (Abdo et al. 2010; Tatischeff \& Hernanz 2007).  That was a symbiotic-like system in which the ejecta had to traverse the wind of the companion red giant during the first week of their expansion and the interaction with the external medium and the collision with the surface of the very extended companion account for the properties of the source.  Instead, here -- and perhaps by extension in other classical novae such as the CO class -- the acceleration is powered by a similar total available kinetic energy but is {\it independent} of the environment.  

Additional support for this scenario comes from the persistent hard X-ray emission during much of the early expansion of these novae.  This, along with the possible presence of very highly ionized species such as [Fe X]  during the NOT sequence for Nova Mon 2012, supports the contention that an ensemble of internal shocks within the ejecta can both heat and accelerate ions to significant energies.   {\bf Indeed, early detections of the [Fe X] 6376\AA\ line in many novae may be due to shock -- rather than photoionization -- excitation and ionization.}   A fragmented structure of the ejecta is found in virtually all novae, manifested in the complexity of the resulting line profiles when observed at high signal to noise and spectral resolution.  This  supports the idea that the actual launching of the expanding matter is far from coherent although the explosion occurs in a short interval of time.  It could be this fact, possibly resulting from the inhomogeneities during the mixing of $\beta$-decaying radionuclides, that produces the low filling factors while accounting for the right order of magnitude in the ejected mass.  In the case of the recurrent novae, e.g. U Sco or T Pyx, the mass of the ejecta may be insufficient to produce detectable emission for any but the closest objects.  Instead, the ONe novae, being the most energetic of the explosions, may be detectable in $\gamma$-rays over a larger Galactic volume even if the optical source is not observable because of interstellar extinction.  For now, we postpone such a quantitative analysis to a future paper.

The other basic result of our study is {\it the ubiquity of the signature for bipolarity (or non-sphericity) of the ejecta}.  This affects the mass estimates from the line fluxes and also points back to the site of the explosion.   This is not a new observation for individual sources, the asymmetries of the profiles have been noted in almost every study as part of the description of the profiles from the time of McLaughlin (1943) and Payne-Gaposchkin (1957).  But the importance of this nearly uniform behavior has been less widely appreciated and it will require a new approach to the modeling of the ignition and ejection mechanism that includes the inherently three dimensional, perhaps four including time, nature of the TNR.

    \begin{acknowledgements}

We have made extensive use of the Astrophysics Data System (ADS), SIMBAD (CDS), and the Barbara A. Mikulski MAST archive at STScI during this work.   The NOT observations were obtained in Fast Track proposal 46-408.  We thank the CHIRON observer, Rodrigo Hernandez, for his hard work.  The STIS spectra were obtained in program DD/TO 13120.  We thank the director of STScI, Matt Mountain, for granting Director's Discretionary Time for the STIS observation and Claus Leitherer, Nolan Walborn,  and Patricia Royle for their help in planning the observations.  We express our deepest thanks to Ronaldo Bellazzini, Neil Gehrels, Elizabeth Hays, Julie McEnery, Peter Michelson, and David Thompson of the {\it Fermi} project for their strong support of the {\it HST} observations and their encouragement, and Olivier Chesneau, Pierre Jean, Jordi Jos\'e, Elena Mason, Ulisse Munari, Jan-Uwe Ness, Kim Page, Soebur Razzaque,  Bob Williams, and the ARAS group for discussions, and Mark Wagner for digging out the V1974 Cyg optical spectra.   C.C.C. was supported at NRL by a Karles' Fellowship and NASA DPR S-15633-Y.  S.S. is grateful for partial support to ASU from NSF and NASA.   We especially thank te referee, Nye Evans, for his careful and supportive review of the original submission.
 \end{acknowledgements}

\newpage
\appendix

\section{Interstellar absorption line measurements}
    \begin{center}
    Table 2.  Interstellar lines in the STIS and NOT spectra\\
    \begin{tabular}{llllll}
    \hline
    Ion & Wavelength$^a$ & gf & EW(m\AA) & SNR \\
    \hline  
Mg II & 1239.93 & 0.000621 & 50$\pm$16 & 3.2\\
Mg II & 1240.40 & 0.000351 & 32$\pm$ 12 & 3.2\\
Si IV & 1402.77 & 0.255 & 129$\pm$36 & 3.6\\
C IV & 1548.20 & 0.190 & 132$\pm$41& 3.2\\
C IV & 1550.77 & 0.0952 &173$\pm$45& 3/8\\
Fe II &1608.45 & 0.0591 & 201$\pm$65& 3.1\\
Fe II & 2586.65 & 0.00717 & 435$\pm$94& 4.2\\
Mn II& 2594.51 & 0.261& 173$\pm$43& 4.2\\
Fe II & 2600.17 & 0.239 & 301$\pm$55& 5.2\\
Mn II & 2606.47 & 0.2 & 156$\pm$35 & 4.4\\
Mg II & 2796.35 & 0.608 & 1054$\pm$32 & 30 \\
Mg II & 2803.53 & 0.303 & 990$\pm$32 & 30\\
Mg I & 2852.96 & 1.800 & 336$\pm$116 & 3\\
\hline
Ca II & 3933.66 & 0.682 & 403$\pm$20\% & (b) \\
Ca II & 3968.47 & 0.330 & 216$\pm$10\%&  (c)\\
Na I  & 5889.95 & 0.641 & 557$\pm$10\% & (b) \\
Na I  & 5895.92 & 0.320 & 428$\pm$10\% & (b)\\
\hline
    \end{tabular}
    \end{center}
 Notes: (a) Wavelengths $>$3000\AA\ are $\lambda_{\rm air}$; (b) Measurements from NOT spectrum Oct. 8; (c) Measurement from NOT spectrum Nov. 21.

\begin{figure}
   \centering
   \includegraphics[width=9cm]{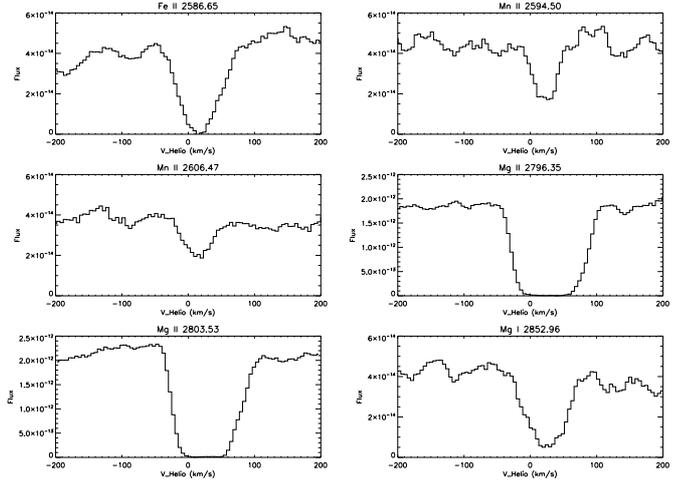}
   \caption{Interstellar absorption line profiles from the STIS spectra, corresponding to those measured for equivalent width. Note the broader velocity range than that covered in the Na I D lines (see Fig. 12).  All velocities are heliocentric.}
    \end{figure}

         \section{Photoionization models: input parameters, measured fluxes, and predictions}  
    \subsection{Measured fluxes}
    
    \begin{center}
    Table 3a. Measured$^{(a)}$ integrated  line fluxes, 2012 Nov. 20-21
    \begin{tabular}{lr}
\hline
Identification & Integrated Flux \\
(ion and wavelength) & erg s$^{-1}$cm$^{-2}$ \\
\hline
N V 1240 &  9.5e-13 \\
 TOT 1300 &   $\leq$4.3e-14 \\
  C II 1335 &   3.5e-14 \\
   TOT 1400 &   1.50e-12 \\
  N IV$]$ 1486 &   4.71e-12 \\
  C IV 1550 &   1.84e-12 \\
  $[$Ne V$]$   1575 &   0.08e-13 \\
   $[$Ne IV$]$   1601 &   1.94e-12 \\
 He II 1640 &   1.06e-12 \\
   O III$]$ TOT   1667 &   2.24e-12 \\
   N III$]$  1750 &   6.6e-12 \\
  C III$]$  1910 &   1.0e-12 \\
 O III     2320 &   2.87e-13 \\
  $[$Ne IV$]$   2423 &   5.21e-13 \\
      2510 &   6.81e-13 \\
      2594 &   6.29e-13 \\
      2630 &   1.58e-12 \\
      2672 &   1.48e-12 \\
   Mg II TOT  2800 &   4.26e-11 \\
   $[$F III$]$  2930 &   1.86e-12 \\
    Ne IV$]$  2974 &   2.27e-12 \\
    O III  3050 & 1.1e-12 \\
\hline    
   $[$Ne III$]$  3869 &   6.79e-11 \\
  $[$Ne III$]$ 3967 &   2.20e-11 \\
     H$\delta$   4101 &   9.11e-12 \\
   He II   4192 &   5.50e-15 \\
   He I   4471 &   1.40e-12 \\
    N III  4520 &   9.97e-13 \\
    H$\beta$  4861 &   2.50-11 \\
    $[$O III$]$ 4959 &   2.28e-11 \\
   $[$O III$]$ 5007 &   7.26e-11 \\
   $[$Fe VII$]$   5177 &   2.39e-12 \\
    O VI  5290 &   1.67e-12 \\
    He II 5411 & 6.81e-13 \\
    $[$N II$]$  5755 &   5.30e-12 \\
  He I    5875 &   7.42e-12 \\
   $[$Fe VII$]$   6087 &   2.78e-12 \\
   $[$O I$]$  6300 &   2.66-12 \\
    $[$Fe X$]$  6476 &   1.48e-12\\
    H$\alpha$  6563 &   1.62e-10 \\
  He I   6678 &   1.57e-12 \\
   He I  7065 &   8.34e-12  \\
   $[$O II$]$  7319 &   3.19e-11 \\
   \hline\end{tabular}
\end{center}    

Note: (a) No extinction corrections have been applied to these data

\begin{figure}
   \centering
   \includegraphics[width=9cm]{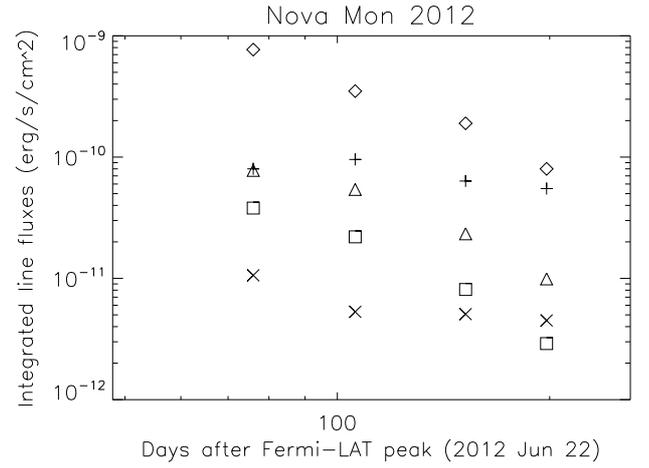}
   \caption{Variation of emission line fluxes with date after outburst (see sec. 2): H$\alpha$ (diamond); H$\beta$ (triangle); [Ne III] 3869\AA\ (plus); [N II] 5755\AA\ (cross); He I 5876\AA\ (square).  The last date is from Munari (2012).  Fluxes are not corrected for E(B-V).}    \end{figure}
 
    \begin{center}
    Table 3b. Time variation of integrated  line fluxes $^{(a)}$
    \begin{tabular}{lllll}
\hline
Line & Day  76 & Day 106 & Day 152 & Day 198$^{(b)}$ \\
\hline
H$\alpha$ & 7.7E-10 & 3.5E-10 & 1.9E-10 & 6.7E-11 \\
H$\beta$  & 7.8E-11 & 5.4E-11 & 2.3E-11 & 9.9E-12 \\
\hline
$[$Ne III$]$ 3869 & 8.5E-11 & 9.6E-11 & 6.4E-11 & 5.5E-11 \\
$[$N II$]$ 5755 & 1.1E-11 & 5.3E-12 & 5.1E-12 & 4.5E-12 \\
$[$O I$]$ 6300 & 6.4E-11 & 4.8E-11 & 2.1E-12 & 1.2E-12 \\
$[$O II$]$ 7320 & 3.1E-11 & $>$1.3E-11$^{(c)}$ & 1.1E-11 & 6.5E-12 \\
\hline
He I 4471 & 5.8E-12 & 2.6E-12 & 1.0E-12 & -- \\
He I 5876 & 3.8E-11 & 2.2E-11 & 8.1E-12 & 2.9E-12 \\
He I 7065 & 3.0E-11 & 1.9E-11 & 7.3E-12 & -- \\
He II 5411 & $<$1E-13 & 1.2E-12 & 6.8E-13 & 3E-13 \\
  \hline
  \end{tabular}
\end{center}    
Notes: (a) Not corrected for reddening, units are erg s$^{-1}$cm$^{-2}$; (b) Measurements for the last date are from Munari (2012, ATel 4709), the others ae from the NOT spectra (Table 1) ; (c) The [O II] 7320\AA\ was heavily absorbed by atmospheric lines on this date, the value quoted is a lower limit.

    \subsection{Photoionization model}

    \begin{center}
Table 3c. Fits of best $Cloudy$ model to dereddend spectrum\\
\begin{tabular}{lcrrrr}
\hline
Ion & Wavelength (\AA)  & F$_{obs}^{(a)}$ & F$_{model}^{(a)}$ & $\log L $ & $\chi^2$ \\
       \hline
TOTL & 1240 &   7.20 &  13.93 & 36.65 &   3.5 \\
C  2 & 1335 &   0.27 &   0.01 & 33.63 &   3.6 \\
TOTL & 1402 &   3.50 &   2.30 & 35.87 &   0.5 \\
TOTL & 1486 &   7.50 &   7.45 & 36.38 &   0.0 \\
TOTL & 1549 &   2.50 &   2.47 & 35.90 &   0.0 \\
Ne 5 & 1575 &   0.50 &   0.90 & 35.46 &   2.5 \\
Ne 4 & 1602 &   2.30 &   4.81 & 36.19 &   4.8 \\
He 2 & 1640 &   1.30 &   5.47 & 36.24 &  41.1 \\
TOTL & 1665 &   2.10 &   0.25 & 34.91 &   3.1 \\
TOTL & 1750 &   6.60 &   0.30 & 34.98 &   3.6 \\
TOTL & 1909 &   1.70 &   0.14 & 34.66 &   3.4 \\
O  3 & 2321 &   0.07 &   0.09 & 34.49 &   0.5 \\
Ne 4 & 2423 &   0.50 &   0.00 &  0.00 &   4.0 \\
Mg 2 & 2803 &  11.30 &   0.01 & 33.69 &   4.0 \\
Ne 3 & 3869 &   6.20 &   5.59 & 36.25 &   0.0 \\
Ne 3 & 3968 &   1.90 &   1.68 & 35.73 &   0.1 \\
H  1 & 4102 &   0.68 &   0.28 & 34.96 &   1.4 \\
H  1 & 4340 &   0.50 &   0.49 & 35.20 &   0.0 \\
TOTL & 4363 &   1.80 &   0.42 & 35.13 &   2.3 \\
He 1 & 4471 &   0.08 &   0.02 & 33.78 &   2.3 \\
H  1 & 4861 &   1.00 &   1.00 & 35.51 &   0.0 \\
O  3 & 4959 &   0.83 &   1.21 & 35.59 &   0.9 \\
O  3 & 5007 &   2.60 &   3.65 & 36.07 &   0.7 \\
Fe 6 & 5177 &   0.08 &   0.09 & 34.45 &   0.1 \\
N  2 & 5755 &   0.12 &   0.00 & 32.12 &   4.0 \\
He 1 & 5876 &   0.17 &   0.05 & 34.23 &   1.9 \\
Fe 7 & 6087 &   0.06 &   0.22 & 34.85 &  32.0 \\
O  1 & 6300 &   0.05 &   0.00 & 27.35 &   4.0 \\
H  1 & 6563 &   2.80 &   3.00 & 35.98 &   0.0 \\
He 1 & 7065 &   0.12 &   0.05 & 34.16 &   1.6 \\
\hline
\end{tabular}
\end{center}
Note: (a) Normalized to dereddened H$\beta$ flux, $4.01\times10^{-10}$ erg s$^{-1}$cm$^{-2}$ for E(B-V)=0.85 (see sec. 4.2).

The irradiating source was assumed to be a blackbody spectrum for this test with the effetive temperature T$_{BB}$ = 3$\times$10$^5$ K, and luminosity of 1$\times$10$^{38}$ erg s$^{-1}$.  For the ejecta, the assumed parameters derived from the line profile modeling were Hydrogen density = 3.2$\times$10$^7$ cm$^{-3}$, Inner radius = 1.5$\times$10$^{15}$ cm, 
Outer radius = 6.0$\times$10$^{15}$ cm, Filling factor = 0.3, and Covering factor = 0.8.  The abundance set, relative to H was:
He =-1.0, C = -3.61, N = -2.52, O = -2.34, Ne = -2.32, Mg = -3.70, Fe = -4.49.  The total $\chi^2/dof$=4.2.  The model was computed without iteration,  fixing the abundances to be those from the V1974 Cyg analysis by Vanlandingham et al. (2006).  

   \section{Comparisons of high resolution ultraviolet spectra of the ONe novae observed at late times: The Fabulous Four}
  
\begin{figure*}
   \centering
   \includegraphics[width=14cm,angle=0]{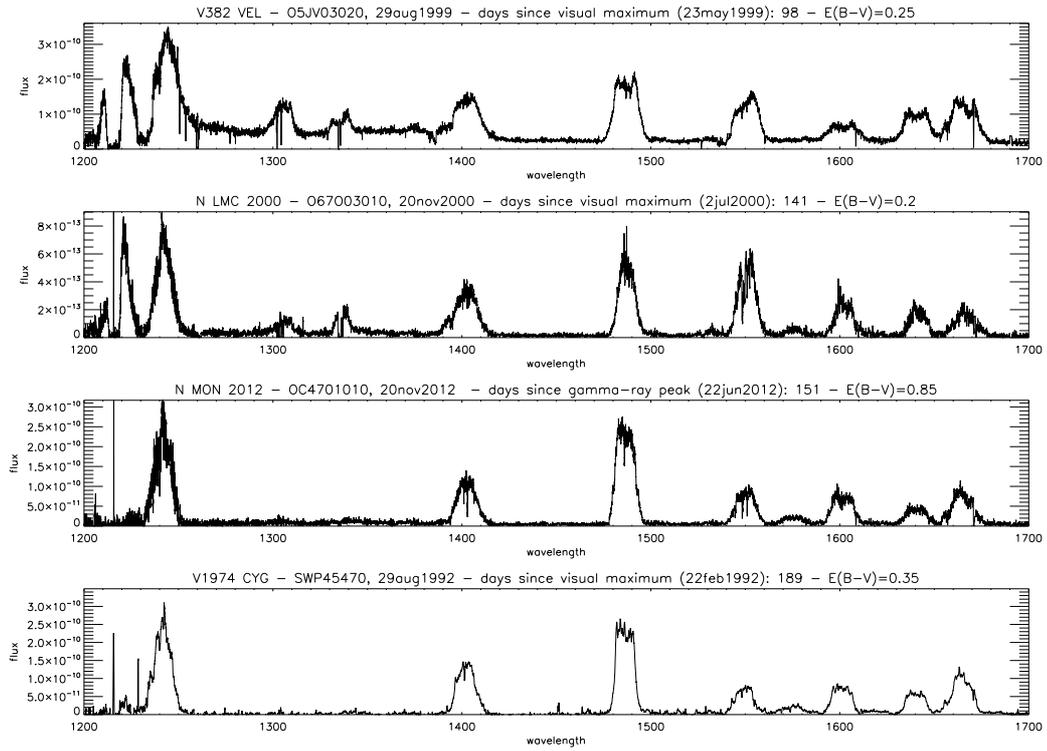}
   \caption{Short wavelength (1200-1700\AA) late time spectra of ONe novae observed at high resolution with {\it IUE} and {\it HST} (GHRS and STIS); time after maximum is noted.  See sec. 4 for discussion.  Note that the designation of epoch for Nova Mon 2012 in both figures is relative to the Fermi/LAT observation and not the visual maximum. }
    \end{figure*}
    
\begin{figure*}
   \centering
   \includegraphics[width=14cm,angle=0]{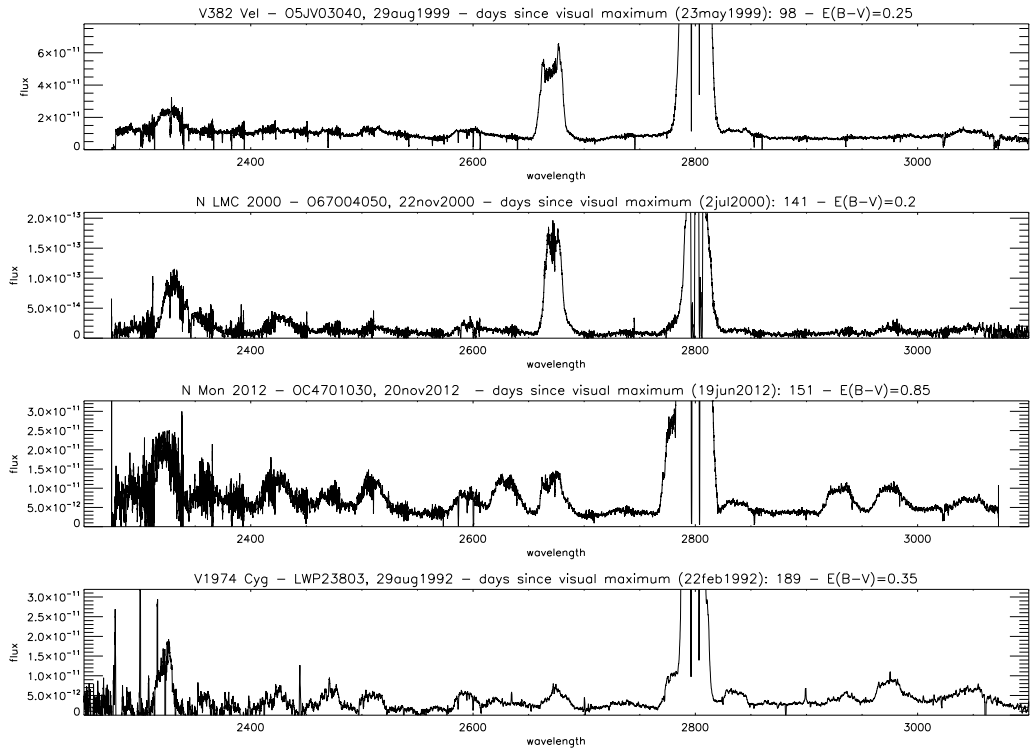}
   \caption{Short wavelength (2300-3100\AA) late time spectra of ONe novae observed at high resolution with {\it IUE} and {\it HST} (GHRS and STIS); time after maximum is noted.  The Mg II 2800\AA\  doublet has been truncated to highlight the weaker lines.}
    \end{figure*}

\end{document}